\documentclass[aip,jcp,reprint]{revtex4-2}
\usepackage{graphicx}
\usepackage{amsmath}
\usepackage{amssymb}
\usepackage{lineno}

\begin{document}

\title{Closed analytical form of many-body free volume and thermodynamics of
monodisperse hard disks}
\author{Victor M. Pergamenshchik}
\affiliation{Institute of Physics of the National Academy of Sciences of Ukraine, Prospekt Nauky, 46, Kyiv 03039, Ukraine}
\affiliation{Center for Theoretical Physics, Polish Academy of Sciences, Al. Lotnik\'{o}w 32/46, 02-668, Warsaw, Poland}
\author{Taras Bryk}
\affiliation{Yukhnovskii Institute for Condensed Matter Physics of the National Academy of Sciences of Ukraine, 1 Svientsitskii Str., Lviv 79011, Ukraine}
\affiliation{Institute of Applied Mathematics and Fundamental Sciences, Lviv National Polytechnic University, 79013 Lviv, Ukraine}
\author{Andrij Trokhymchuk}
\altaffiliation{Corresponding author at: adt@icmp.lviv.ua}
\affiliation{Yukhnovskii Institute for Condensed Matter Physics of the National Academy of Sciences of Ukraine, 1 Svientsitskii Str., Lviv 79011, Ukraine}
\affiliation{Faculty of Chemistry and Chemical Technology, University of Ljubljana, Ve\v{c}na pot 113, 1000 Ljubljana, Slovenia}

\begin{abstract}
The hard-disk model is a fundamental reference system for excluded-volume
physics, liquid structure, and two-dimensional ordering. Yet, despite its
apparent simplicity, an analytical formulation of its thermodynamics has
remained difficult because the free volume available to disks' centers has a
highly nontrivial many-body geometry. Here we show that this geometry admits
a closed and exact representation: for any given hard-disk configuration,
the free volume can be analytically expressed through intersection areas of
up to five exclusion disks. This provides a direct geometrical route from
particle coordinates to the configurational partition function and entropy.
We prove that the $N$-disk partition function factorizes into a product of
conditional free volumes and identify its two limiting asymptotic forms: the
low-density fluid regime controlled by extensive cavities and the
high-density fluid regime controlled by intensive private cells. Using the
geometric measures computed from representative equilibrium particle
configurations, the resulting theory reproduces the established hard-disk
equation of state over almost the entire density range up to close packing.
At intermediate densities, the analysis reveals a mixed-fluid regime in
which localized caged defects provide an additional configurational entropy {%
within the density interval} preceding liquid-hexatic coexistence. Finally,
the five-disk intersection area is shown to be a sensitive local scalar
measure of hexagonal ordering, complementing the orientational order
parameter and emphasizing the central role of many-body correlations.
\end{abstract}

\maketitle

\affiliation{Institute of Physics of the National Academy of Sciences of
Ukraine, Prospekt Nauky, 46, Kyiv 03039, Ukraine} \affiliation{Center for
Theoretical Physics, Polish Academy of Sciences, Al. Lotnik\'{o}w 32/46,
02-668, Warsaw, Poland}

\affiliation{Yukhnovskii Institute for Condensed Matter Physics of the
National Academy of Sciences of Ukraine, 1 Svientsitskii Str., Lviv 79011,
Ukraine} \affiliation{Institute of Applied Mathematics and Fundamental
Sciences, Lviv National Polytechnic University, 79013 Lviv, Ukraine}

\altaffiliation{Corresponding author at: adt@icmp.lviv.ua} %
\affiliation{Yukhnovskii Institute for Condensed Matter Physics of the
National Academy of Sciences of Ukraine, 1 Svientsitskii Str., Lviv 79011,
Ukraine} \affiliation{Faculty of Chemistry and Chemical Technology,
University of Ljubljana, Ve\v{c}na pot 113, 1000 Ljubljana, Slovenia}

\section{Introduction}

For nearly two centuries since van der Waals proposed his equation of state,
hard spheres and hard disks have served as the quintessential reference
models for simple liquids, glasses, colloids, and numerous other condensed
matter systems\cite{ref6,SpatialStat}. The two-dimensional (2D) hard-disk
(HD) system is the molecular physics counterpart of the 2D Ising model in
magnetism; yet, unlike the 2D Ising model, analytical thermodynamic results
for the HD fluid remain exceptionally scarce \cite{ref14}. Under these
circumstances, the numerical Monte Carlo and molecular dynamics approaches
have become the main tools in the study of 2D HD systems (see reviews \cite%
{ref14,ref14a} and numerous references therein). In particular, since Alder
and Wainwright's 1962 discovery of the hard disk phase transition\cite{ref8}%
, the complete equation of state has been predominantly obtained through
extensive numerical experiments\cite%
{ref14,ref14a,ref11,ref12,ref13,ref9,ref10}.

The sequence of ten coefficients of the virial expansion for the HD model 
\cite{virial10} decreases very slowly indicating that at best convergence of
the whole series is very slow. Remaining nevertheless a tool used for
fitting the HD equation of state, it yields an incorrect behavior already at
densities above $\sim $0.6 (i.e., far from the close packing density $\sim
0.907$). The primary problem in the theory of hard-core systems is the lack
of a small interaction parameter\cite{Feynman}. The potential energy of a HD
system is zero and the entropy, which in this case is determined by a purely
geometric structure of the phase space, gives the total thermodynamic
information as a function of density which is the single parameter of the
system state. It has long been recognized that the entropy of a hard-core
system should be directly related to the so-called free volume -- the area
accessible to a disk's center given the coordinates of all other disks~\cite%
{ref2,ref3,ref15,ref17,Hoover72,ref19,ref20,Speedy80,ref21,Speedy91,Stillinger92,ref22,ref24,2014,2015,2016,2023,Isobe}%
. However, the complex shape and connectedness of the 2D free space make its
precise determination and measurement difficult\cite{ref31,ref32}.
Historically, theoretical and numerical access to the free volume relied on
Widom's insertion method associated with the extensive fraction of free
volume, the cavity. In an $N$-disk system, this is related with the
probability of inserting into an equilibrium system~of ($N-1$) disks an
additional $N$-th disk\cite{Widom}. However, at moderate to high densities,
finding a finite cavity to accommodate a new particle becomes a task futile~%
\cite{ref3,ref24} thus resulting in a paradox: $N$-th disk of an $N$-disk
system has no place therein. At the same time, {Hoover et al} 
%Ashurst, and Groover 
\cite{Hoover72,ref19} have pointed out that the free volume also has an
intensive fraction, a private cell related to a single disk, which is
nonzero even if cavity vanishes. 
%\textbf{In the present work we distinguish between the conventional local free volume associated with a single particle and the global free volume entering the configurational partition function.}
This, however, has not resolved the insertion problem as the extensive and
intensive free volume fractions were addressed only on separate grounds and
an exact determination of the free volume remains a problem. Moreover, the
unsolved problems are also a relation between the free volume and the
partition function (PF) and the very method which could adequately capture
the complex geometric structure of the free volume. In this situation, hard
spheres and HDs have been addressed by very different approaches. The
essential connection with hard disks in the modern glass theory has
motivated development of the mean field theory for hard spheres in high
nonphysical dimensions, from which one can apparently descend to the 2D and
3D \cite{ref4,ref5}.

In recent decades, pioneering works by Tanaka and co-workers have
fundamentally reshaped our understanding of liquid-state physics,
demonstrating that simple two-body density correlations are inherently
insufficient to capture the onset of structural ordering and fluid
instability \cite{tanaka2022roles,leocmach2013importance}. Instead,
many-body angular correlations and medium-range crystalline order play the
defining role in structural transitions and dynamic heterogeneity,
particularly in 2D hard-core systems \cite{kawasaki2007correlation,
russo2015assessing}. In line with this profound insight, our geometric
approach rigorously incorporates many-body correlations by evaluating the
simultaneous intersection areas of up to five disks thereby capturing the
structural coherency that two-body insertions miss \cite{tanaka2022roles}.

The paper is organized as follows. In Section~II, we first overview the
previous related development over past decades, II.A, and then formulate the
statistical geometry of the free volume and derive formulae for both its
extensive and intensive components in terms of multi-disk intersection
areas. 
%\textbf{Section~II introduces the analytical framework. Subsection~II.A derives the exact geometrical formulation of the global free volume together with its decomposition into the extensive cavity and intensive private-cell contributions, whereas Subsection~II.B presents the analytical evaluation of the intersection measures $\mu_2$, $\mu_3$, $\mu_4$, and $\mu_5$ that determine the free volume.}
In Section~III, we show that the configurational partition function
factorizes into a product of conditional free volumes and establish its
limiting asymptotic forms in the low-density fluid (dubbed "gas-like") and
high-density fluid (dubbed "liquid-like") regimes. In Section~IV, we compute
the equation of state and identify the crossover from low-density fluid
regime to high-density fluid regime followed by the mixed-fluid regime in
which the system gains a defect-induced entropy. In Section~V, we introduce
the five-disk intersection area as a local scalar indicator of the hexagonal
ordering. {Finally, in Section~VI we summarize the results and discuss their
implications and possible extensions. The technical details of the
analytical geometry of multi-disk intersections, the close-packing limit,
the derivation of the partition function and simulation details are provided
in Appendices~\ref{app:intersections}, \ref{app:lattice}, \ref{app:partition}
and \ref{app:simulations}. }

\section{Statistical Geometry of the Free Volume}

\subsection{Free volume, cavity, and equations of state: the previous
development.}

\begin{figure}[t]
\centering
\includegraphics[width=\columnwidth]{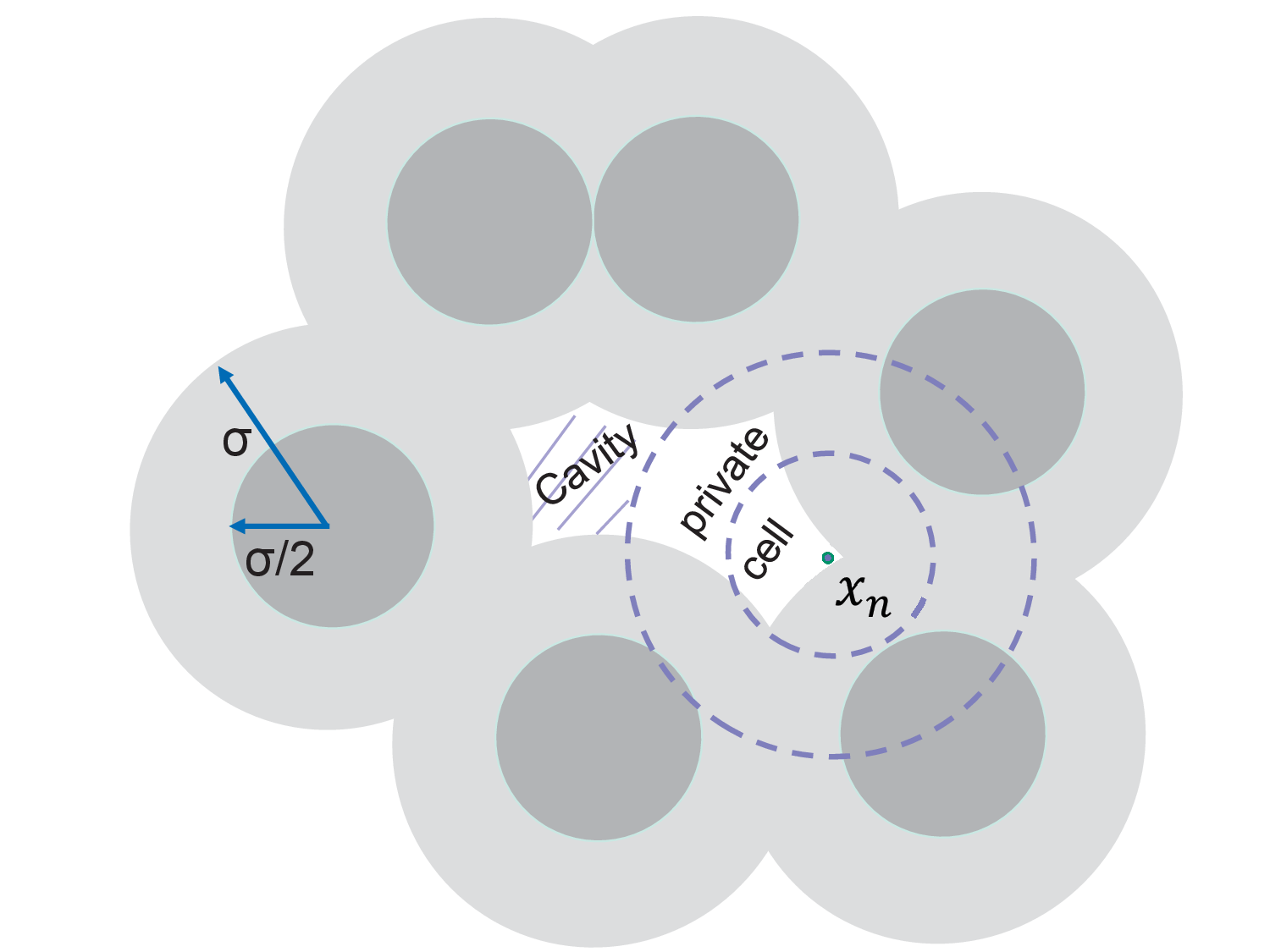}
\caption{Fragment of an equilibrium system of $N$ hard disks. The ($N-1$)
hard disks are dark circles, and the connected $\protect\sigma$-circles are
light circles. The $n$-th disk and circle are shown by dashes. The inner
white area is the {global} free volume of the $n$-th disk, the hatched
fraction is its cavity, and the clear fraction is its private cell.}
\label{fig:1}
\end{figure}

Here we give a brief review of the main ideas and quantities involved in the
theory of HD model and their development in the past decades. We start with
presenting these quantities in terms of the tools which we introduced and
use in our approach. A hard disk has radius $\sigma/2$, the hard core. To
determine the free volume, we assign a concentric \textquotedblleft $\sigma$%
-circle\textquotedblright\ of radius $\sigma$ to each disk. No disk center $%
x_{n}$ can enter $\sigma$-circles of other disks so that their union forms
an excluded volume. While the cores cannot overlap, $\sigma$-circles
themselves can overlap without violating the hard-core repulsion. A
configuration of $N$ hard disks in a 2D volume $V$ consists of $N$
non-overlapping cores and $N$ connected $\sigma$-circles which can overlap,
Fig.~\ref{fig:1}. Depending on the approach and context, the free volume and
cavity, introduced to address the area of $V$ not covered by the exclusion
circles, can be defined as global or as local quantities. 
%\textbf{Because our approach is formulated directly in terms of the configurational partition function, we consider this free area as a global quantity}
As in Ref.~\cite{ref20} our approach is based on the PF where this free area
appears as a global quantity which can comprise more than one separate
connected areas (in Fig.~\ref{fig:1}, a single one is shown). Consequently,
we define the free volume $V_{N,n}$ of a disk\ centered at $x_{n}\in V$ in a
given configuration of a $N$ disk system as the area $V_{N,n}$ accessible
for its center in this configuration. This is the total $V$ minus union of
the rest $(N-1)$ $\sigma $-circles: 
\begin{equation}
V_{N,n}=V-\mu \left( \cup _{i\neq n}^{N}\sigma _{i}\right) ,  \label{1.1}
\end{equation}%
where $i=1,2,...,N$ is disk's number and $\mu $ is the area. The cavity in a
HD system of $N$ disks is the total area of $V$ where an additional disk,
call it $N+1$-st disk, can be inserted. To connect to this definition, we
denote cavity in a $N$ disk system as $\left\langle C_{N+1}\right\rangle_{N}$%
. It is total $V$ minus union of all $N$\ $\sigma $-circles: 
\begin{equation}
\left\langle C_{N+1}\right\rangle _{N}=V-\mu \left( \cup _{i=1}^{N}\sigma
_{i}\right) \,.  \label{2.1}
\end{equation}%
It is these quantities (\ref{1.1}) and (\ref{2.1}) that naturally appear in
the expression for the PF (\ref{ZNN}), which 
%\textbf{naturally leads to the definitions of the global free volume and global cavity adopted in the present formulation.}
justifies our definitions of the free volume and cavity as global quantities
in the total $V.$ 
%\textbf{The distinction between the global and local definitions therefore concerns the decomposition of the accessible space into connected components rather than the underlying excluded-volume geometry.}
Although in the global approach $V_{N,n}$ and $\left\langle
C_{N+1}\right\rangle _{N}$ formally equal to sums of their values for all
spatially separated connected areas, {the individual connected regions need
not be determined explicitly.} 
%actually these local contributions are of no need for their computation. 
As will be seen in the next Section, this is an essential simplification as
the global $V_{N,n}$ and $\left\langle C_{N+1}\right\rangle _{N}$ are fully
determined just by the $N$ disk coordinates. At the same time, the
approaches of Refs.\cite{Hoover72,ref19,Speedy80,ref21,Speedy91,ref24} 
%\textbf{were formulated in terms of individual connected free regions.}
required to consider local connected free areas $v$ individually.
Consequently, these papers define the free volume and cavity for each
connected area $v$ in a way equivalent to using the formulae similar
respectively to (\ref{1.1}) and (\ref{2.1}) (in~\cite{ref24} the union of
local cavities equal to the global cavity $\left\langle C_{N+1}\right\rangle
_{N}$ was called available space). This implies restricting $x_{n}$ to $v$, $%
x_{n}\in v$, which is not a trivial task as it requires first to determine
the boundary of the connected area $v$ and identify disks therein. In spite
of this difference, free volume and cavity in both global and local
approaches share %\textbf{common geometrical properties.}
certain general common properties. If the free volume is $V$ or $v$ without
the area covered by $(k-1)$ $\sigma $-circles then cavity is $V$ or $v$
without the area covered by $k$ $\sigma$-circles, i.e., by one more, which
shows that the free volume in both $V$ and $v$ is larger than or equal to
the cavity, in particular, $V_{N,n}\geq \left\langle C_{N+1}\right\rangle
_{N}.$ The difference between the free volume and cavity was first
articulated by Hoover and coworkers \cite{ref9,Hoover72,ref19} who argued
that the extensive cavity volume which is "space available for a new
particle throughout the system", is smaller than "space available for a
particle which is already existing in the system". This difference is a
space which is intensive, scales as $V/N,$ and consists of individual,
private single-disk cells. 
%\textbf{The free volume available for a particle already present in the system naturally corresponds to the global free volume defined by Eq.~(\ref{1.1})}
The free volume available for a particle which is already in the system
exactly corresponds to our definition (\ref{1.1}) and the above inequality:
it is the sum of the cavity and private cell, and when the cavity is getting
smaller and smaller, the total free volume reduces to the volume of
individual cells. This is illustrated in Fig.~\ref{fig:1}: the private cell $%
c_{N,n}$ of an $n$-th disk (white clear) in a HD system of $N$\ disks is the
free volume of this disk $V_{N,n}$ (white) without the cavity $\left\langle
C_{N+1}\right\rangle_{N}$ (hatched): 
\begin{equation}
c_{N,n}=V_{N,n}-\left\langle C_{N+1}\right\rangle _{N}.  \label{3.1}
\end{equation}
%
%\textbf{Equation (\ref{3.1}) shows that the global free volume naturally decomposes into an extensive contribution (the cavity) and an intensive contribution (the private cell), thereby providing a unified description of both limiting concepts used in the previous literature.}
A triangular lattice provides an exact example: the free volume vanishes
only at the close packing fraction $\eta_{\mathrm{cp}}\simeq 0.907$ whereas
cavity vanishes already at $\eta \simeq 0.3$ (Appendix~\ref{app:lattice}).

%\begin{figure}[htpb]

\begin{figure}[t]
\centering
\includegraphics[width=\columnwidth]{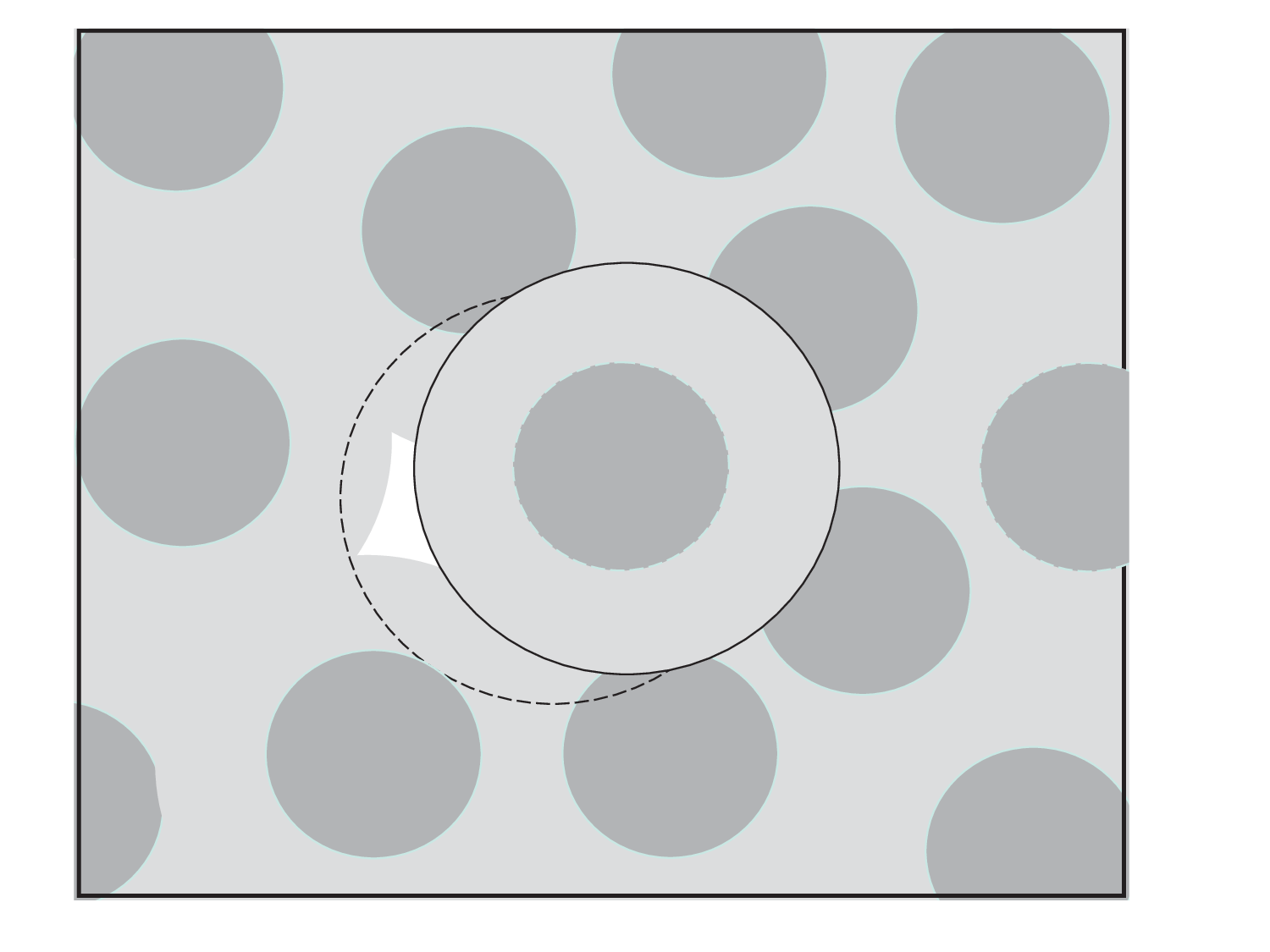}
\caption{A configuration of ($N-1$) hard disks in an equilibrium $N$-disk
system; the $N$-th disk is transparent, not seen, and its center is in the
white area. The shown ($N-1$) disk configuration is allowed as its $\protect%
\sigma $-circles leave a cell (white area) for the $N$-th disk. However, it
is different from the equilibrium configuration of ($N-1$) disks which is
homogeneous. In the equilibrium ($N-1)$ system one of the $N-1$ disks, whose 
$\protect\sigma $-circle is indicated by solid line, would be at the
position shown by dashes. This dashed $\protect\sigma $-circle fully covers
the white area so that another, $N$-th disk, cannot be inserted. This is a
disallowed configuration as the $N$-th disk has no place in it. Thus,
insertion of $N$-th disk in the equilibrium system of $(N-1)$ disks, as
suggested by Widom's approach~\protect\cite{Widom}, is impossible whereas in
the $N$-disk system the $N$-th disk has the finite free volume shown by the
white area. }
\label{fig:7}
\end{figure}

Based on this idea, in 1972, Hoover, Ashurst, and Groover~\cite{Hoover72}
showed that the pressure can be expressed via the inverse ratio of the free
volume to its surface (perimeter in 2D) \cite{Hoover72}. After that, in a
series of papers \cite{ref20,Speedy80,ref21,Speedy91} Speedy introduced the
spare volume and defined it as a quantity determined by \textquotedblleft
...the probability that another sphere can be placed at a random point in
the assembly\textquotedblright\ \cite{ref20}. {In the present notation,} 
%Clearly, 
this is equivalent to a global cavity available for an additional, $N+1$-st
disk in a system of $N$ disks. Later Speedy explicitly shifted from the
nomenclature of spare volume to cavity \cite{Speedy80,ref21,Speedy91}. As a
result, using the approximation of discrete lattice, Speedy related the PF
of a HD system with the product of cavities $\left\langle C_{k}\right\rangle
_{k-1}$ in the systems of $k$ disks, but defined via a disk insertion in a
system of $(k-1)$ disks \cite{ref20,Speedy80,Speedy91}: 
\begin{equation}
Z_{SW}=\left\langle C_{N}\right\rangle _{N-1}\left\langle
C_{N-1}\right\rangle _{N-2}...\left\langle C_{k}\right\rangle _{k-1}...V.
\label{ZN2}
\end{equation}%
Developing this idea, Speedy proposed a formula similar to that by Hoover
and co-workers \cite{Hoover72} which however expressed the HD pressure via
the inverse ratio of the average local cavity to its average perimeter. As
total cavity consists of multiple connected areas, Speedy redefined cavity
as a single local connected area available for the insertion. Since both
cavities and perimeters are additive, this definition did not eliminate the
essential {difficulty of the} insertion {approach:} 
%this definition could not eliminate the essential problem related to the underlying insertion idea: 
In paper~\cite{Speedy91} the authors expressed concern that at high
densities cavities can vanish, but preferred to proceed without any
amendment. 
%\textbf{Consequently,} Speedy's formulation based on the PF (\ref{ZN2}) \textbf{is primarily applicable in the} low-density 
Thus, at best Speedy's formula for the pressure and the PF (\ref{ZN2}) can
address a low density range in which cavities dominate and private cells are
negligible.

The connection of Speedy's results to cavities and in particular the way the
PF (\ref{ZN2}) was obtained were going back to the earlier results by Adams 
\cite{Adams} and Andrews \cite{Andrews} which had in turn been inspired by
Widom's approach \cite{Widom}. Widom's idea \cite{Widom} was to replace the
integral over $k$ coordinates in the PF of $N$ particles by the PF $Z_{k}$
of an \textit{equilibrium} system of $k$ particles. This is an approximation
valid only for a sufficiently low fluid density. Indeed, whereas in the
exact integration configurations of $k$ particles are subject to the
condition that the rest $(N-k)$ particles are accommodated in the system's
volume, in the equilibrium $k$ particle system the integration is over all
their configurations no matter if they leave place for the rest $(N-k)$
particles, which assumes that the system is large and $N$ is small. Thus the
PF $Z_{SW}$ may be referred to as %can rightfully be called 
Speedy-Widom partition function.

This result along with Eq.~(\ref{1.1}) is behind the insertion method, which
has been intensively used in the study of hard-disk and hard-sphere systems.
Equation~(\ref{1.1}) expresses the free volume via the union of just $(N-1)$ 
$\sigma $-circles, which has prompted the simple 
%\textbf{but generally invalid identification} that
but generally unjustified idea that the free volume $V_{k,k}$ of the $k$-th
disk in a $k$-particle system can be computed as a cavity $\left\langle
C_{k}\right\rangle _{k-1}$ in the \textit{equilibrium} system of $(k-1)$
disks. This identification ignores the fact that whereas such configuration
is homogeneous and can leave no void for an extra disk, in a $k$-disk system
the configuration of $(k-1)$ $\sigma$-circles forms the exclusion union that
has a void for $k$-th disk. This is illustrated in Fig.~\ref{fig:7}. As a
result, already for densities well below freezing, finding cavities in
computer simulations was dubbed a task futile \cite{ref24}. This situation
is a paradox: resorting to the insertion idea, we started to study a system
of disks and found that one of them has no space in this system. The reason
is that, in general, $V_{k,k}\geq\left\langle C_{k}\right\rangle _{k-1}$,
with equality only in the low-density limit. 
%and the equality is only the low-density approximation. 
As shown in the next subsection, Eq.~(\ref{1.1}) can be identically
rewritten in the form which refers only to the original $N$-disk system and
resolves the paradox.

%This identification ignores the fact that whereas \textbf{the equilibrium $(k-1)$-disk configuration can be homogeneous and leave no void for an additional disk, the configuration of $(k-1)$ exclusion circles occurring within the original $k$-disk system must leave accessible space for the $k$-th disk.} This is illustrated in Fig.~\ref{fig:7}. As a result, already at densities well below freezing, finding cavities in computer simulations was dubbed a task futile \cite{ref24}. The resulting paradox is that insertion into an independently equilibrated $(k-1)$-disk system can fail even though the $k$-th disk has a finite free volume in the original $k$-disk system. The reason is that, in general, 
%$V_{k,k}\ge \langle C_k\rangle_{k-1}$, with equality only in the low-density limit.}

A central difficulty in the theory of HD model and, in particular, for
implementing the formulae proposed by Hoover, Ashurst, and Groover~\cite%
{Hoover72} and Speedy~\cite{Speedy80} is that exact formulae for the free
volume, cavity, and their perimeters have not been known. 
%{A central difficulty in} the theory of the HD model, and in implementing the formulae proposed by Hoover, Ashurst, and Groover~\cite{Hoover72} and Speedy~\cite{Speedy80}, is that %\textbf{explicit analytical expressions} for the free volume, cavity, and their perimeters have not been available.
The numerical simulations consist of producing independent configurations of
disks' coordinates which is the task input, and then certain related theory
must provide calculation of the free volume, entropy, equation of state, and
other quantities of interest which are the task outcome. In principle, the
ultimate theory must give the outcome directly from the disks' coordinates,
but actually the available theoretical methods relate disks' coordinates
with the free volume, cavities, etc. only through intermediate quite
sophisticated geometrical and semi-empirical constructions such as lattice
approximation \cite{Stillinger92} and polynomial extrapolation \cite{ref22},
Voronoi cells and Delaunay simplices \cite{ref22,ref24,2014}, sophisticated
algorithms \cite{2015,Isobe} and just advanced insertion procedures \cite%
{2016,2023}, to name a few. 
%\textbf{provide equilibrium configurations of disk coordinates. A theoretical framework is then required to connect these coordinates to the free volume, entropy, equation of state, and other thermodynamic quantities. Ideally, this connection should be as direct as possible. Existing approaches typically involve intermediate} geometrical and semi-empirical constructions, including lattice approximations~\cite{Stillinger92}, polynomial extrapolations~\cite{ref22}, Voronoi cells and  Delaunay simplices~\cite{ref22,ref24,2014}, specialized geometric algorithms~\cite{2015}, or insertion procedures~\cite{2016,2023}.
In the next subsection the ultimate problem of the free volume, cavity, and
private cell is solved: we derive exact analytical formulae relating these
quantities with the disks' coordinates. 
%\textbf{we derive explicit analytical expressions for} the free volume, cavity, and private cell \textbf{directly in terms of} the disk coordinates.

\subsection{Exact analytical expressions for the global free volume,
extensive cavity, and intensive private cell.}

We consider an equilibrium system of $N$ hard disks of radius $\sigma /2$
contained within an area $V$. The concentric $\sigma $-circles introduced in
the previous subsection and their intersections play the main role in our
approach as the union of $\sigma $-circles is the excluded volume where
centers of other disks cannot enter. Equation~(\ref{1.1}) for the {global}
free volume is exact, but it is written explicitly in terms of $(N-1)$ disks
while the $N$-th disk labeled by $n$ appears only implicitly, which can
prompt an idea of connection with a system of $(N-1)$ disks. To avoid this
we rewrite formula for $V_{N,n}$ in an identical form which refers only to
the original $N$-disk system. This form is \cite{CMP2023} 
\begin{eqnarray}
V_{N,n} &=&V-\mu \left( \cup _{i=1}^{N}\sigma _{i}\right) +\pi \sigma
^{2}-\mu \left[ \sigma _{n}\cap (\cup _{i\neq n}^{N}\sigma _{i})\right] 
\notag \\
&=&\left\langle C_{N+1}\right\rangle _{N}+\pi \sigma ^{2}-\mu \left[ \sigma
_{n}\cap (\cup _{i\neq n}^{N}\sigma _{i})\right]\,.  \label{Vn}
\end{eqnarray}
Its meaning is clarified in Fig.~\ref{fig:1}: the $\left\langle
C_{N+1}\right\rangle _{N}$ is the cavity, Eq.~(\ref{2.1}) (hatched), and $%
\pi \sigma ^{2}-\mu \left[\sigma_{n}\cap (\cup_{i\neq n}^{N}\sigma _{i})%
\right] $ is area of the $\sigma $-circle of disk $n$ without the area of
its overlapping $\sigma_{n}\cap (\cup_{i\neq n}^{N}\sigma_{i})$ with all
other (gray) $\sigma $-circles. 
%But this is the clear white area, i.e., the private cell $c_{N,n}$ of disk $n$. 
{This area is precisely the clear white region in Fig.~\ref{fig:1}}, i.e.,
the private cell $c_{N,n}$ of disk $n$. Thus, the private cell and the
global free volume are 
%Thus, the private cell and the total free volume are
%
\begin{eqnarray}
c_{N,n} &=&\pi \sigma ^{2}-\mu \left[ \sigma _{n}\cap (\cup _{i\neq
n}^{N}\sigma _{i})\right] ,  \label{cNn} \\
V_{N,n} &=&\left\langle C_{N+1}\right\rangle _{N}+c_{N,n},  \label{C + c}
\end{eqnarray}%
which agrees with Eq~(\ref{3.1}). We see that both $\left\langle
C_{N+1}\right\rangle _{N}$ and $c_{N,n}$ are expressed in terms of the $N$
disks in the original system and can therefore be computed if the
coordinates of $N$ disks are known. 
%No other system needs to be involved and no insertion is needed at all. 
{Thus, both quantities are determined entirely by the coordinates of the
original $N$-disk configuration; no auxiliary system or particle-insertion
procedure is required.}

The intensive private cell $c_{N,n}$ is the {particle-centered contribution
to the global} free volume of disk $n$, equal to the fraction of its own {$%
\sigma$-circle not overlapped by the other exclusion circles;} 
%The intensive private cell $c_{N,n}$ is the free volume of disk $n$ equal to the fraction of its own area which is notoverlapped with other $\sigma $-circles; 
the $c_{N,n}$ is obviously intensive and does not grow with the system size.
In contrast, the cavity $V-\mu \left(\cup_{i=1}^{N}\sigma_{i}\right)$ is
extensive and scales with the total size $V$. The expressions for the cavity
and private cell as well as Fig.~\ref{fig:1} show that $\left\langle
C_{N+1}\right\rangle _{N}$ is not related to any specific disk but $c_{N,n}$
of a disk $n$ depends on its coordinate $x_{n}$. For this reason the 
%total 
{global} free volume of disk $n$ also depends on $x_{n}$ which is indicated
by the subscripts $n$ in Eq.~(\ref{C + c}). This dependence has an intensive
character and disappears on averaging over disks in a given configuration.

Using the inclusion-exclusion formula of the set theory, both $\left\langle
C_{N+1}\right\rangle _{N}$ and $c_{N,n}$ can be exactly expressed in terms
of the intersection areas of the $\sigma $-circles. As the hard cores cannot
overlap, a maximum of five $\sigma $-circles can intersect at any given
point (to clarify this fact, see Fig.~\ref{fig:9} and its caption). The
intersection area of two $\sigma $-circles, centered at $x_{n}$ and $x_{i}$,
is the area $\mu _{ni}$ common to both $\sigma $'s; similarly, $\mu
_{nij},\mu _{nijk},$ and $\mu _{nijkl}$ are the areas common for three,
four, and five $\sigma $-circles, respectively. Now we introduce the
following disk intersection areas $\mu _{k,n}$ involving disk $n:$ $\mu
_{2,n}=\sum_{i}^{^{\prime }}\mu _{ni},$ $\mu _{3,n}=\sum_{i>j}^{^{\prime
}}\mu _{nij},$ $\mu _{4,n}=\sum_{i>j>k}^{^{\prime }}\mu _{nijk},$ $\mu
_{5,n}=\sum_{i>j>k>l}^{^{\prime }}\mu _{nijkl},$ where prime indicates that
all summation indices are different from $n$. In terms of $\mu _{k,n}$, the
second term in (\ref{cNn}) is given by the known formula of the set theory,
and one has 
\begin{equation}
c_{N,n}=\pi \sigma ^{2}-(\mu _{2,n}-\mu _{3,n}+\mu _{4,n}-\mu _{5,n}).
\label{c1}
\end{equation}
Using the same formula, the second term in (\ref{2.1}) can be presented as
the sum of all intersection areas $\mu _{k,n}$ of each $\sigma _{n}$-circle
with all other $\sigma $'s, regarding the fact that, in this sum, the
intersections of each two, three, four, and five $\sigma $-circles appear
respectively two, three, four, and five times. Then 
\begin{eqnarray}
\left\langle C_{N+1}\right\rangle _{N} &=&V-\left[ N\pi \sigma ^{2}-\right.
\\
&&\left. \sum_{n=1}^{N}\left( \frac{\mu _{2,n}}{2}-\frac{\mu _{3,n}}{3}+%
\frac{\mu _{4,n}}{4}-\frac{\mu _{5,n}}{5}\right) \right] \,.  \notag
\label{C1}
\end{eqnarray}%
Of course, $\left\langle C_{N+1}\right\rangle _{N}$, $c_{N,n},$ and $V_{N,n}$
depend on a specific configuration $\{x\}_{N}$ of $N$ disks. We see that $%
\left\langle C_{N+1}\right\rangle _{N}$ consists of the sums over disks'
coordinates in the state $\{x\}_{N}$ which suggests that it can be exactly
expressed in terms of the intersection areas averaged over all $N$ disks in
the configuration $\{x\}_{N}$. We introduce the mean values $\mu _{k}$ in a
given single state $\{x\}_{N}$ : 
\begin{equation}
\mu _{k}=\frac{1}{N}\sum_{n=1}^{N}\mu _{k,n}\quad \mathrm{for}\quad
k=2,3,4,5\,.  \label{muk}
\end{equation}%
In terms of these mean intersection areas, the cavity, private cell, and
free volume themselves acquire the status of mean values in a single given
configuration $\{x\}_{N}$, and we denote them $\left\langle
C_{N+1}\right\rangle _{N},\left\langle c_{N}\right\rangle $, and $%
\left\langle V_{N}\right\rangle $, respectively (Pergamenshchik \cite%
{CMP2023}): 
\begin{eqnarray}
\left\langle C_{N+1}\right\rangle _{N} &=&V-N\left[ \pi \sigma ^{2}-\right. 
\notag \\
&&\left. \left( \frac{\mu _{2}}{2}-\frac{\mu _{3}}{3}+\frac{\mu _{4}}{4}-%
\frac{\mu _{5}}{5}\right) \right] ,  \label{CN} \\
\left\langle c_{N}\right\rangle &=&\pi \sigma ^{2}-(\mu _{2}-\mu _{3}+\mu
_{4}-\mu _{5}),  \label{cN} \\
\left\langle V_{N}\right\rangle &=&\left\langle C_{N+1}\right\rangle
_{N}+\left\langle c_{N}\right\rangle .  \label{Vaverage}
\end{eqnarray}

The exact analytical formulae for computing the intersection areas $\mu _{2}$%
, $\mu _{3}$, $\mu _{4}$, and $\mu _{5}$ from particle coordinates are
presented in Appendix~\ref{app:intersections}. Notably, at the disk
close-packing, $\eta _{\mathrm{cp}}=\pi /(2\sqrt{3})\approx 0.907$ (packing
fraction $\eta =\pi \sigma ^{2}N/4V$), both extensive cavity and intensive
cell identically vanish, $C_{\mathrm{cp}}=c_{\mathrm{cp}}=0$, validating the
functional independence and precision of these geometric definitions
(explicitly demonstrated in Appendix~\ref{app:lattice}). 
%\textbf{This can be used to provide an explicit example of a configuration in which there is no cavity but the free volume is nonzero. In a perfect triangular lattice the cavity (\ref{2.1}) vanishes already at $\eta\simeq 0.3$,  whereas the private cell is finite for $0.3<\eta<0.907$.}

\section{Factorization of the Partition Function}

One of the main problems of statistical mechanics is to calculate the
partition function \cite{BaxterReview}. In the thermodynamic limit and even
in a system of large number of particles, PF is practically an
infinite-dimensional integral which cannot be computed directly even
numerically. The problem of statistical mechanics is to solve this integral
which implies to somehow reduce its calculation to a finite number of some
other actions \cite{Baxter}. This is trivial in the case of ideal gas as
then computing the PF reduces to that of a finite-dimensional integral. The
HD model is not of this kind and we have only the second choice. In this
section, we show that, via the free volume, the PF of the HD model can be
expressed in terms of just four functions of density, i.e., $\mu _{k}$ (\ref%
{muk}), which can be computed by the above analytical formulae using disks'
coordinates \cite{Analyticaltheory}.

In this section we show that the four intersection areas not only determine
the free volume, but also the PF and thus the entropy of the hard-disk
system. The reason is that the multidimensional configurational PF $Z$
factorizes into a product of conditional free volumes resembling the
factorial expression and can be computed by sequential integrations (details
are given in Appendix \ref{app:partition}). The configurational PF of $N$
hard disks in the volume $V$ is the integral over accessible disks'
coordinates:%
\begin{equation}
Z\propto \int_{V}dx_{1}...dx_{N}\exp \left( -\sum_{i<j}^{N}U_{ij}\right) \,,
\label{Z}
\end{equation}%
where the interaction between disks $i$ and $j$ of the radius $\sigma /2$ is 
$U_{ij}=\infty $ for $x_{j}$ within $\sigma _{i}$ and $U_{ij}=0$ for $x_{j}$
outside $\sigma _{i}.$ Up to a factor specified below, $Z$ can be rewritten
in the following highly suggestive form: 
\begin{widetext}
\begin{eqnarray}
Z&=&\int_{V}dx_{1}\int_{V}dx_{2}e^{-U_{21}}\int_{V}dx_{3}e^{-U_{31}-U_{32}}...\int_{V}dx_{k}e^{-
\sum_{j=1}^{k-1}U_{kj}}...\int_{V}dx_{N}e^{-\sum_{j=1}^{N-1}U_{Nj}}
\label{ZNN} 
 \\
&=&\int_{V}dx_{1}\int_{V_{2}\{x_{1}\}_{N}}dx_{2}\int_{V_{3}\{x_{1},x_{2}\}_{N}}dx_{3}...\int_{V_{k}\{x_{k-1} 
\}_{N}}dx_{k}...\int_{V_{N}\{x_{N-1}\}_{N}}dx_{N}\,.  
%\label{GPF} 
\notag
\end{eqnarray}
\end{widetext}The structure of the integration domains makes this integral
in a certain sense similar to a factorial. Indeed, formally, the integration
domain in $x_{1}$ consists of all coordinates from $V$ such that at the same
time the rest ($N-1$) disks can be somehow (i.e., their actual arrangement
is irrelevant) accommodated in $V$. The integration domain $%
V_{2}\{x_{1}\}_{N}$ in $x_{2}$ consists of all coordinates $x_{2}$ from $V$
given disk $1$ is at $x_{1}$ and at the same time the rest ($N-2$) disks can
be somehow accommodated in $V$. Similarly (Appendix \ref{app:partition}),
the integration domain $V_{k}\{x_{k-1}\}_{N}$ of disk $k$ in an $N$-disk
system is the fraction of $V$ accessible to the center of $k$-th disk for
all possible rearrangements of $N-k$ disks in the maze made by a given
allowed configuration $\{x_{k-1}\}$ of $k-1$ disks; an allowed configuration 
$\{x_{k-1}\}$ is subject to the condition that at the same time the rest ($%
N-k+1$) disks can be somehow accommodated in $V$. Thus, PF (\ref{ZNN})
factorizes into the product of the mean conditional free volumes computed
over accessible configurations of $k$ disks in $V$: 
\begin{equation}
Z\propto \prod_{k=1}^{N}\langle V_{k}\rangle _{N}\,.  \label{ZZ}
\end{equation}

%\begin{figure}[htpb]
\begin{figure}[t]
\centering
\includegraphics[width=\columnwidth]{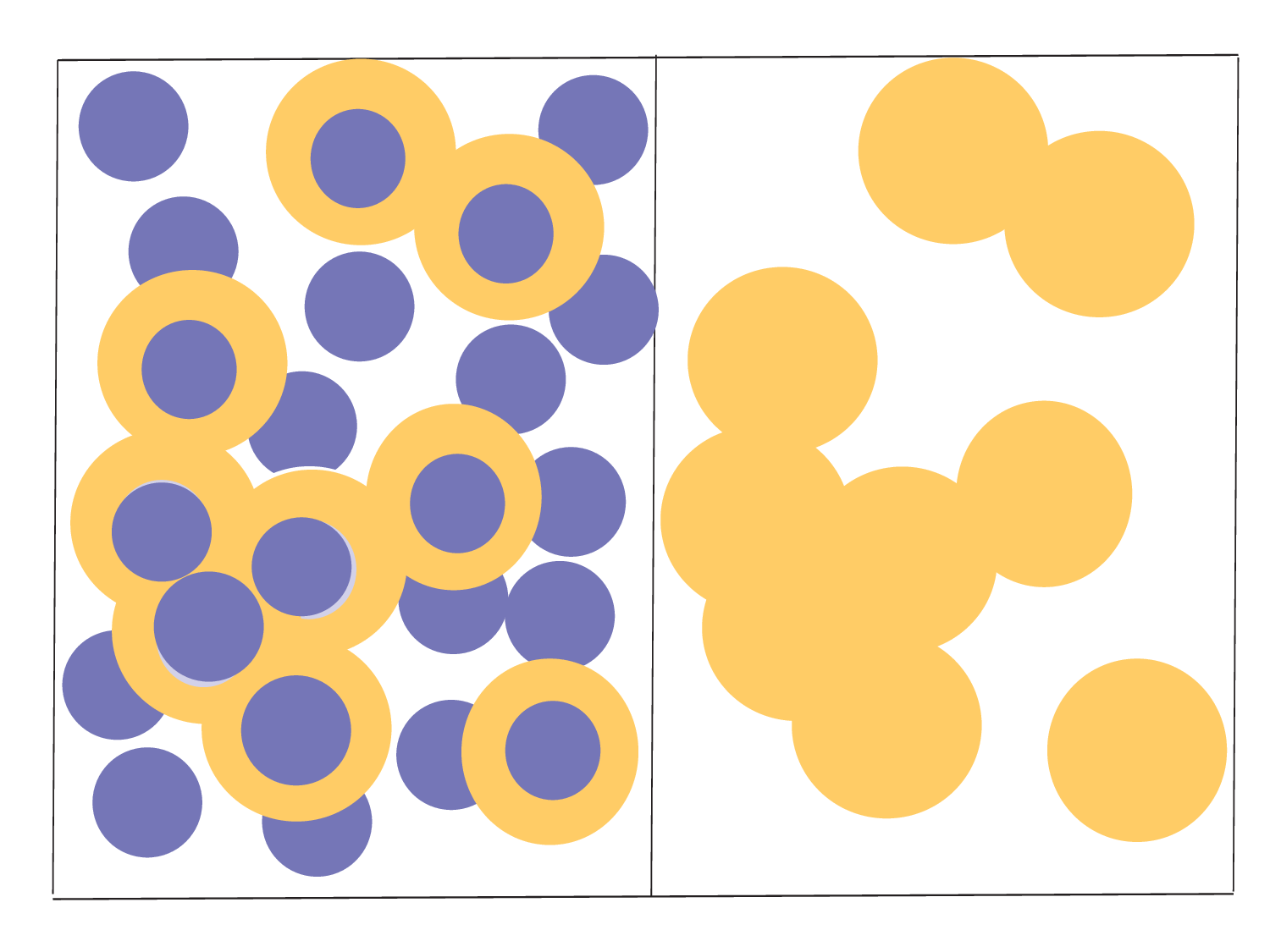}
\caption{Left panel: A system of $N=23$ hard disks with dark cores. The
chosen $k-1=9$ hard disks have light concentric $\protect\sigma $-circles.
Right panel: The light $\protect\sigma -$circles are those in the left panel
and the empty area is the free volume $V_{10}\{x_{9}\}_{23}$ for any dark
disk chosen as the 10-th disk. As here $\protect\eta $ is below 0.5, the
rest $N-k+1=14$ hard disks are accommodated for any positions of the chosen
9 disks, the cavity dominates the {global} free volume, and the gas-like
approximation (G) is justified.}
\label{fig:2}
\end{figure}
Actually, the similarity between the PF (\ref{ZZ}) and a factorial is formal
and ends for high densities. In the mathematical factorial, the condition of
accommodation of any number $N-k$ of disks is a priory fulfilled for any
position $x_{k}$ of disk $k$, for any configuration $\{x_{k-1}\},$ as well
as for any $N/V.$ But in a real physical system, moving a disk $k$ from its
position $x_{k}$ to another position $x_{k}^{\prime }$ is possible if other
disks can give a physical space to such a move. Clearly, at low density this
is not a problem, but at high densities this implies a coherent
rearrangement of positions of many other disks, which is statistically
almost impossible in a highly frustrated system. As a result, the PF (\ref%
{ZZ}) has different limiting forms for low and high densities. For a low
density, the conditional free volume and one allowed configuration are
illustrated in Fig.~\ref{fig:2}. An allowed and not allowed configurations
for a high density are shown in Fig.~\ref{fig:7}.

%The conditional free volume $V_{10}\{x_{9}\}_{23}$ of $10$-th disk in a system of 23 disks and one allowed configuration $\{x_{9}\}_{23}$ of 9 disks are illustrated in Fig.~\ref{fig:2} for the case of low density. An allowed and not allowed configurations for a high-density are shown in Fig.~\ref{fig:7}.

We arrived at the result that the partition function $Z$ has the form of a
factorial-like sequential construction of the accessible $N$-body
configuration space which is built as a product of conditional geometric
measures. Importantly, this representation reveals that the complexity of
the hard-disk problem needs not be represented by an intrinsically infinite
hierarchy of many-body correlations, 
%is not due to an intrinsically infinite hierarchy of interactions, 
but can be reduced to a finite set of many-body geometric correlations
encoded in the free volume. Because the latter is exactly expressible
through intersection areas of up to five disks, the factorized form of the
partition function establishes an exact finite %analytical reduction 
closure of the configurational integral at the level of five-body
correlations. This provides a complementary non-perturbative formulation to
traditional approaches based on virial expansions or integral equation
theories where the many-body problem is treated through infinite series or
approximate closures \cite{ref2,McQuarrie}.

Depending on the packing fraction $\eta =\pi\sigma^{2}N/(4V)$, the exact PF (%
\ref{ZZ}) reduces to the two asymptotically exact limiting forms,
corresponding to a lower and higher density, which we conventionally dub
"gas"-like and "liquid"-like approximation. But before proceeding, to avoid
misunderstanding, it is instructive to outline the meaning of "gas" and
"liquid" in the HD model.

Although there is no mathematical proof excluding a gas-liquid transition in
the HD model, it is generally accepted that the configurational PF is
analytic up to the fluid-hexatic transition near 
%Although so far there is no mathematical proof of non-existence of a gas-liquid transition in the HD model, it is likely that the PF is analytic up to 
$\eta \simeq 0.7$ so that in this range of $\eta $ the low-density state
smoothly transitions to higher-density states \cite{ref14}. The phase
sequence in a HD system has been established by advanced MC simulations by
Krauth and co-workers\cite{ref11,ref12}, confirmed in experiments with 2D
colloids\cite{ExperColloids}, and was reviewed in Ref.\cite{ref14}.
According to these studies, a disordered fluid is below $\eta \simeq 0.7$ at
which a first-order fluid-hexatic transition occurs and the two-phase
coexistence in the narrow density range from $\eta\approx 0.700$ to $\approx
0.716$ takes place. 
%The hexatic-solid transition is continuous and occurs at $\eta\approx 0.720$.
Above $\eta \simeq 0.716$ up to $\simeq 0.720$, the phase is hexatic with a
quasi-long hexagonal orientational and no translational order, and at $\eta
\simeq 0.720$ it continuously transitions to a solid with a quasi-long range
translational and long range orientational order. At last at the close
packing the order becomes crystalline. This phase sequence is in line with
the Kosterlitz-Thouless-Halperin-Nelson-Young scenario\cite{KTHNY} except
that the phase transition to the hexatic phase at $\eta \simeq 0.7$ is found
to be not continuous \cite{ref11,ref12}.

Let us now proceed with the general PF (\ref{ZNN}). Although the HD model
possesses neither distinct gas and liquid phases nor a gas-liquid phase
transition, the low-density and high-density asymptotic representations 
%Although gas and liquid are indistinguishable in a HD system, the low-density and high-density asymptotics 
of the PF are different. For this reason, it is convenient to associate
these two asymptotic representations with the descriptive labels "gas-like"
(G) and "liquid-like" (L). These terms refer exclusively to the limiting
asymptotic forms of the PF and should not be interpreted as denoting
distinct thermodynamic phases of the hard-disk system. 
%So it is instructive to be connected to the specific forms of the PF. 
We dub the G approximation the PF form $Z_{G}$ (\ref{ZG}) because it in
particular includes very low densities that correspond to the dilute-fluid
limit. %can be naturally associated with a gas. 
The high-density regime %The density range 
$\eta >0.58$ contains both disordered fluid, hexatic and solid phases, but
we dub it the L approximation to underline that PF $Z_{L}$ (\ref{ZL})
corresponds to the high-density fluid limit. 
%Depending on the packing fraction $\eta =\pi \sigma ^{2}N/(4V)$, the exact form of partition function (\ref{ZZ}) resolves into two distinct asymptotic limits governed by the actual dominant system's configurations:

\begin{enumerate}
\item \textbf{The G approximation:} At low densities ($\eta \lesssim 0.53$),
the empty space is large, comparable with its occupied complement and thus
highly connected. Then all $N$ disks can be accommodated in $V$ for any
positions of both ($k-1$) disks and disk $k$ (see Fig.~\ref{fig:2}).
Effectively, the $k$-th disk interacts only with the chosen $(k-1)$ disks in
determining the accessible configurational space. 
%One may say that the $k$th disk sees only the chosen ($k-1$) disks. 
In the connected area disks can swap positions during a single relaxation
time (see Appendix~\ref{app:partition}.3). In this case the calculation of
the PF is standard: the integration is formally extended over all
indistinguishable configurations which is then compensated by the dividing
by $N!$\,. Thus, in the low-density or G regime the cavity fraction
dominates the free volume, yielding the expression 
\begin{equation}
Z_{\mathrm{G}}=\frac{1}{N!}\prod_{k=1}^{N}\langle C_{k}\rangle _{k-1},
\label{ZG}
\end{equation}%
where $\langle C_{k}\rangle _{k-1}$ is the cavity of a disk in a system of
just $k-1\simeq k$ disks, i.e., which has the density $\rho ^{\prime }=k/V$.
Then the entropy $S_{\mathrm{G}}=\ln Z_{\mathrm{G}}$ can be converted to the
integral%
\begin{equation}
S_{\mathrm{G}}=V\int_{0}^{N/V}d\rho ^{\prime }\ln \left\langle C(\rho
^{\prime })\right\rangle -N\ln N/e\,.  \label{S G}
\end{equation}

%\begin{figure}[htpb]
\begin{figure}[t]
\centering
\includegraphics[width=\columnwidth]{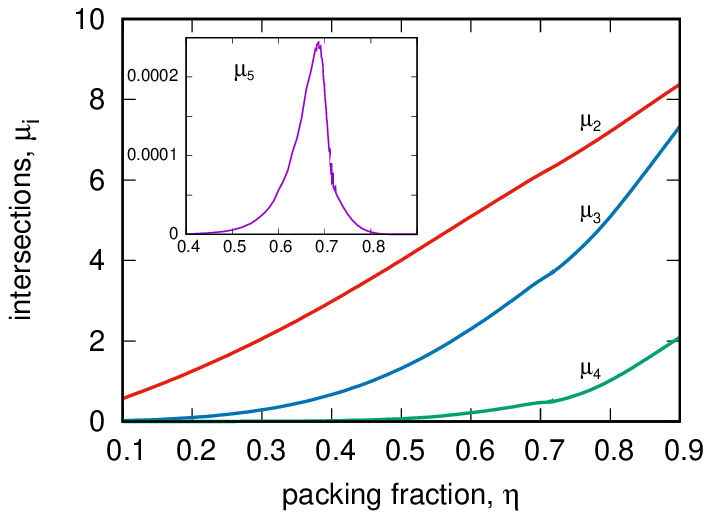}
\caption{The mean intersection areas $\protect\mu_{2}$, $\protect\mu_{3}$,
and $\protect\mu_{4}$ of two, three, and four $\protect\sigma$-circles as
functions of the packing fraction $\protect\eta $. Inset: the mean
intersection area $\protect\mu _{5}$ of five $\protect\sigma$-circles. The
diameter is set to $\protect\sigma =1$\,.}
\label{fig:3}
\end{figure}

\item \textbf{The L approximation:} At higher densities ($\eta \gtrsim 0.58$%
), local cavities disconnect and disks are persistently confined by the
cages formed by their neighbors (see Fig. \ref{fig:7}). As was stressed in
subsection II.A, in the dominating disks' configurations, cavities become
extremely rare, a position swapping is also practically an impossible event, 
%\textbf{and the cavity contribution to the global free volume becomes negligible} in the thermodynamic limit
i.e., both are fluctuations which can be neglected in the thermodynamic
limit (see discussion in Appendix~\ref{app:partition}.3). As each disk can
explore only its private cell $c$, all $V_{k}$ tend to the mean common value 
$\langle c_{N}\rangle$, 
%the combinatorial factor $1/N!$ is no longer required because particle permutations no longer generate statistically distinct equilibrium configurations.
the division by $N!$ has no sense as disks do not exchange their positions.
Thus, in what we will refer to as the L regime, the PF $Z_{\mathrm{L}}$ and
the entropy $S_{\mathrm{L}}$ crossover into 
\begin{eqnarray}
Z_{\mathrm{L}} &=&\langle c_{N}\rangle ^{N},  \label{ZL} \\
S_{\mathrm{L}} &=&N\ln \langle c_{N}\rangle .  \notag
\end{eqnarray}
\end{enumerate}

\begin{figure}[t]
\centering
\includegraphics[width=\columnwidth]{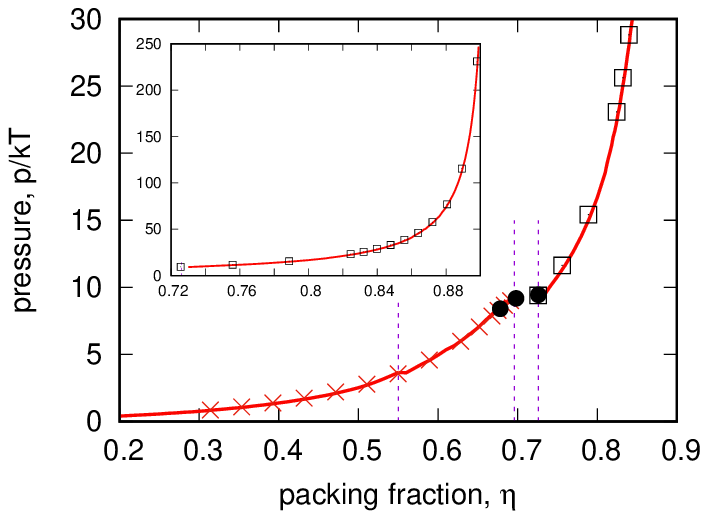}
\caption{The pressure $P$ predicted by theory (solid line) and superimposed
on the simulation data $P_{\mathrm{exp}}$ (crosses\protect\cite{ref10},
empty squares\protect\cite{ref9} and filled circles\protect\cite%
{ref14,ref11,ref12}). The vertical dash lines delimit, from left to right,
the G region, the crossover (with a small mismatch slightly rightward of $%
\protect\eta =0.55$) and the mixed-fluid regions, the fluid-solid phase
coexistence region $0.696<\protect\eta<0.720$ (empty), and the L region.}
\label{fig:4}
\end{figure}

The distinction between the G and L regimes has a direct geometrical and
dynamical interpretation. At low packing fractions, the intersection areas
of two $\sigma$-circles dominate (see Fig. \ref{fig:3}), indicating that the
correlations of more than two particles are negligible. As a result, the
free volume is dominated by large cavities. In large cavities particles
explore the configuration space extensively and exchange positions over the
relaxation time (Appendix~\ref{app:partition}.3). 
%This effectively restores the statistical equivalence of particle permutations
This effectively restores permutation symmetry leading to the standard
factor $1/N!$ in the PF and the pressure diverging at close packing as $\ln
(V_{N})$ (Appendix~\ref{app:partition}.2). In contrast, at high packing
fractions, the intersection areas of three and four $\sigma $-circles and
thus the three- and four-particle correlations are essential. Then the free
volume becomes fragmented into disconnected, particle-centered regions
(private cells). In this regime, particle motion is restricted to local
cages formed by nearest neighbors, and global exploration of configuration
space through particle exchange is dynamically suppressed. As a result, 
%the dominant contribution to the configurational integral no longer arises from particle permutations
the configurational integral no longer samples permutations of particle
coordinates and the PF reduces to a product of local contributions without
the $1/N!$ factor. The broken connectivity of the free volume structure at
high densities ensures that the theoretical pressure diverges precisely as $%
1/\langle c_{N}\rangle$ at close packing, completely mirroring empirical
data (see Fig. \ref{fig:4}). This shows that distinction between lower and
higher densities in a HD system is not related to an energetic competition,
but to the change of connectivity and topology of the accessible
configuration space. This provides a purely geometrical interpretation of
the crossover from delocalized to caged dynamics, consistent with the modern
views of liquid structure and dynamical heterogeneity based on many-body
correlations\cite{kawasaki2007correlation,tanaka2022roles,ref5}.

%\begin{figure}[htpb]

\section{Equation of State and Mixed-fluid Region}

To validate the theoretical entropy and equation of state, we computed the
intersection area functions ($\mu _{2}$ through $\mu _{5}$) {analytically
from the particle coordinates using the exact formulae} 
%derived in Appendix~\ref{app:intersections}.
%exactly from particle coordinates using analytical formulas 
(see Appendix~\ref{app:intersections}). The coordinates were obtained from
representative equilibrium configurations generated via $NVT$ Monte Carlo
simulations of $N=900$ hard disks across the density range $0.1\leq\eta \leq
0.9$ (see Appendix~\ref{app:simulations} for simulation details). 
%{No adjustable parameters enter the evaluation of the intersection measures.}

%Although simulated, we do not address in our analysis the density range $0.69\lesssim \eta \lesssim 0.72$ which includes two-phase coexistence, studied in great detail in \cite{ref11,ref12}. The reason is that proper computing of averages for $\mu_{k}$ in this density interval requires a substantially larger-scale simulations with $N\sim 10^{6}$ disks \cite{ref11,ref12}. Further, from the exact intersection areas, the theoretical pressures $P_{\mathrm{G}}$ and $P_{\mathrm{L}}$ were calculated as functions of the density by the entropy differentiation with respect to $V$: $P_{\mathrm{G,L}}=\partial S_{\mathrm{G,L}}/\partial V$ with $V=N/(4\eta /\pi )$ ($k_{B}T$ is set one).

\begin{figure}[b]
\centering
\includegraphics[angle=-90, width=\columnwidth]{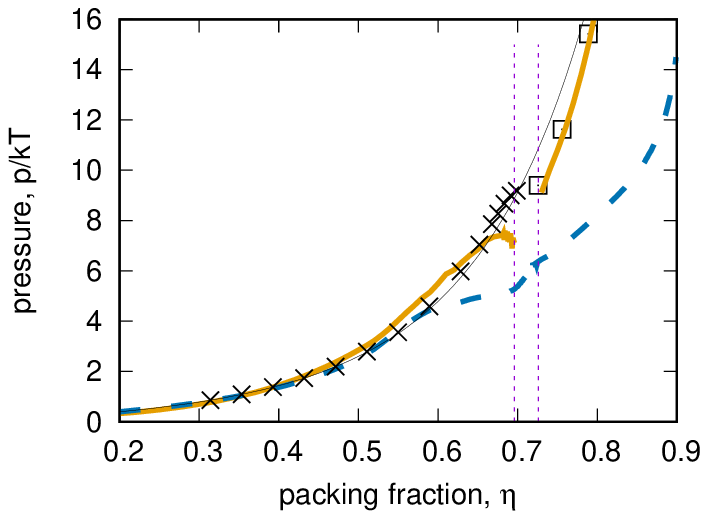}
\caption{The pressures $P_{\mathrm{exp}}$ (crosses\protect\cite{ref10} and
empty squares\protect\cite{ref9}, the same as in Fig.~\protect\ref{fig:4}), $%
P_{\mathrm{L}}$ in the L approximation (thick yellow solid line), $P_{%
\mathrm{G}}$ in the G approximation (thick blue dashed line), and the virial
pressure resulting from the ten virial coefficients \protect\cite{virial10}
(thin solid line).}
\label{fig:5}
\end{figure}

Further, obtained intersection areas were used to calculate the theoretical
pressures $P_{\mathrm{G}}$ and $P_{\mathrm{L}}$ as functions of the density
by the entropy differentiation with respect to $V$: $P_{\mathrm{G,L}%
}=\partial S_{\mathrm{G,L}}/\partial V$ with $V=N/(4\eta/\pi )$ ($k_{B}T$ is
set one). Although simulated, we do not address in our analysis the density
range $0.69\lesssim \eta \lesssim 0.72$ which includes two-phase
coexistence, studied in great detail in \cite{ref11,ref12}. The reason is
that proper computing of averages for $\mu_{k}$ in this density interval
requires a substantially larger-scale simulations with $N\sim 10^{6}$ disks 
\cite{ref11,ref12}. %Although the density range $0.69\leq\eta\leq0.72$
%which consists of two-phase fluid-hexatic coexistence and the hexatic phase, 
%was also simulated, we do not address it since the description of these thermodynamic states requires highly advanced simulations with $N\sim10^6$ disks~\cite{ref11,ref12}. 
The theoretical predictions in remaining density intervals $\eta\lesssim 0.69
$ and $\eta\gtrsim 0.72$ successfully fit the known computer simulation data~%
\cite{ref14,ref10,ref9} for equation of state, $P_{\mathrm{exp}}$,
distinguishing four density regions (see Fig.~\ref{fig:4}):

\begin{itemize}
\item \textbf{G region ($\eta \lesssim 0.53$):} Excellent agreement is
achieved using the purely cavity-driven G approximation limit $P_{\mathrm{G}%
}(\eta)$\,.

\item \textbf{L region ($\eta \gtrsim 0.72$):} Excellent agreement is
achieved using the purely cell-driven L approximation limit $P_{\mathrm{L}%
}(\eta )$. The L approximation accurately describes the thermodynamic
behavior, showing an accurate pressure divergence at $\eta_{\mathrm{cp}}$ .

\item \textbf{Crossover region ($0.53\lesssim \eta \lesssim 0.58$):} The
system transitions from a macroscopically connected free volume to localized
caging.

\item \textbf{Mixed-fluid region ($0.58\lesssim \eta \lesssim 0.68$):}
Figure~\ref{fig:5} shows that for $\eta >$ $0.55$, the pure L approximation
pressure $P_{\mathrm{L}}$ first goes above $P_{\mathrm{exp}}$ indicating
that as $\eta$ increases, the actual entropy decreases slower than the pure
L approximation entropy $S_{\mathrm{L}}$ and hence is higher than $S_{%
\mathrm{L}}$. The additional entropy gain can be found by recognizing that
the system is not perfectly homogeneous, but acts as a mixture of already
caged disk \textquotedblleft defects\textquotedblright\ {associated with the
increasing local hexagonal ordering}, 
%signaling of the approaching hexagonal ordering transition, 
and still uncaged disks.
\end{itemize}

%\begin{figure}[htpb]

The deviation of $P_{\mathrm{L}}$ of the pure L approximation from the
experimental computer simulation data in this intermediate density range
indicates that the {assumption} of a structurally homogeneous system {%
becomes insufficient.} 
%idea of a structurally homogeneous system is no longer sufficient. 
Instead, the underlying configurations exhibit spatial heterogeneity with
regions of strongly caged particles within the disordered fluid environment
at the current density $\eta$. Indeed, analyzing 2D hard-disk
configurations, Huerta et al \cite{ref44} revealed that defects in the form
of hard disks caged by their neighbors appear already at densities around $%
\eta \sim 0.5$, their fraction grows with $\eta $ and tends to 1 at the end
of the phase 
%reaching $\sim 40\% $ and $\sim 60\%$ on the fluid and solid boundaries of the fluid-solid c
coexistence region $0.696<\eta<0.720$. Such defects, constrained to the
local density at the ordering point ($\eta_{\mathrm{d}}\approx 0.68$), {are
consistent with the emergence of local} %indicate the nascent 
hexagonal ordering. To account for this effect, we introduce a
phenomenological defect fraction $\zeta (\eta )$ that represents the
proportion of regions where the local packing approaches the characteristic
density $\eta_{\mathrm{d}}\approx 0.68$. While $\zeta (\eta )$ is not
derived here from the first principles, its introduction is guided by the
observed structural heterogeneity of HD configurations and provides a
minimal extension of the L approximation that captures the additional
configurational entropy.

%\begin{figure}[htpb]
\begin{figure}[t]
\centering
\includegraphics[width=\columnwidth]{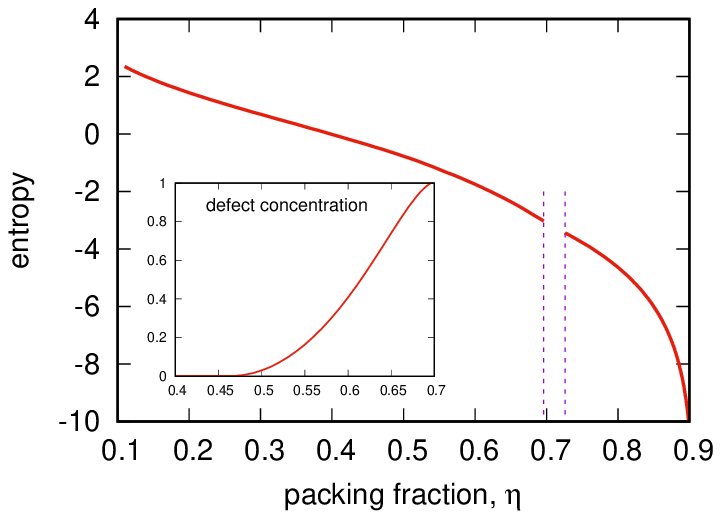}
\caption{The entropy $S$ smoothly interpolating the different thermodynamic
regions as a function of the packing fraction $\protect\eta $. Inset:
phenomenological fraction of caged defects, $\protect\zeta(\protect\eta)$\,. 
}
\label{fig:6}
\end{figure}

Assuming $\zeta N=N_{\mathrm{d}}$ is the number of defects, the partition
function of the hard disks-defect mixture is 
\begin{equation}
Z_{\mathrm{L,mix}}=(N-N_{\mathrm{d}})Z_{\mathrm{L}}(\eta )\times N_{\mathrm{d%
}}Z_{\mathrm{L}}(\eta _{\mathrm{d}})\frac{N!(N-N_{\mathrm{d}})!}{N_{\mathrm{d%
}}!}\,  \label{Zmix}
\end{equation}%
and the corresponding entropy reads: 
\begin{eqnarray}
S_{\mathrm{L,mix}} &=&N[(1-\zeta (\eta ))\ln \left\langle c_{N}(\eta
)\right\rangle +\zeta (\eta )\ln \left\langle c_{N}(\eta _{\mathrm{d}%
})\right\rangle  \notag \\
&&-(1-\zeta (\eta ))\ln (1-\zeta (\eta ))-\zeta (\eta )\ln \zeta (\eta )]\,.
\label{Smix}
\end{eqnarray}
%
%\textbf{In analogy with the Kosterlitz-Thouless scenario\cite{KT1973}, the presence of localized defects contributes to an additional configurational entropy.}
Thus, as in the Kosterlitz-Thouless scenario\cite{KT1973}, generating
defects the system increases its entropy. The complete entropy $S$ of
monodisperse hard disks, providing a smooth interpolation between 
%smoothly interpolating 
the entropies $S_{\mathrm{G}}$, Eq.~(\ref{S G}), the above $S_{\mathrm{L,mix}%
}$, and $S_{\mathrm{L}}$, Eq.~(\ref{ZL}), respectively in the G,
mixed-fluid, and L ranges of $\eta$ is presented in Fig.~\ref{fig:6}. The
pressure $P_{\mathrm{exp}}$ in the mixed-fluid range in Fig.\ref{fig:4} is
well fitted by $\partial S_{\mathrm{L,mix}}/\partial V$ for the defect
fraction $\zeta (\eta )$ shown in the inset of Fig.~\ref{fig:6}. The
smoothness of the entropy curve {provides strong support for} 
%omitting the factor \(1/N!\) in the high-density partition function \(Z_L\), Eq. (19).
%The smoothness of the entropy curve is a strong indication of the correctness of 
omitting the factor $1/N!$ in the expression for the high-density PF $Z_{L}$%
, Eq.~(\ref{ZL}).

Furthermore, the analytically identified mixed-fluid region -- where the
system behaves as a statistical mixture of uncaged disks and localized caged
disk defects -- provides an analytical {framework consistent with} Tanaka's
two-order-parameter model, wherein a liquid is viewed as a dynamic mixture
of normal liquid structures and competing locally favored structures \cite%
{tanaka2022roles}. In 2D hard-core systems, these defect-like structures
geometrically frustrate macroscopic crystallization while simultaneously
promote the emergence of medium-range {local} crystalline order and dynamic
heterogeneity \cite{kawasaki2007correlation}. 
%\textbf{Although the present theory is formulated for hard disks, the resulting picture of coexisting locally favored and less constrained environments closely resembles Tanaka's phenomenology of structural heterogeneity.}
Specifically, a structurally analogous bimodality in the local density,
explicitly distinguishing \textquotedblleft gas-like\textquotedblright\
(uncaged) and \textquotedblleft liquid-like\textquotedblright\ (caged)
states, has recently been observed in extremely confined quasi-1D hard disks
at commensurate widths~\cite{Trokhymchuk2025}. This demonstrates that the
entropically driven local structural heterogeneity, which we treat
analytically via the defect fraction $\zeta (\eta )$, is a fundamental,
measurable physical feature of hard-core fluids influenced by approaching
phase transformations.

\section{Scalar Indicator of the Hexagonal Order}

One of the central outcomes of the present geometrical framework 
%The central outcome of the presented geometrical framework 
is that many-body structural information on the HD system is naturally
encoded in intersection measures $\mu _{k}$, Fig.~\ref{fig:3}. In
particular, we show here that the five-disk intersection area $\mu _{5}$
captures the emergence and quality of the local hexagonal environment.

While intersection areas $\mu _{2},\mu _{3},$ and $\mu _{4}$ increase
monotonically with $\eta $ due to progressive compaction, $\mu _{5}(\eta )$
exhibits fundamentally different behavior (Fig.~\ref{fig:3}, inset). It
rises to a sharp maximum at the onset of the fluid-hexatic phase transition (%
$\eta\sim 0.700$) and then decays to zero at close packing. This distinction
highlights an essential limitation of two-body and low-order correlation
measures: they primarily encode packing constraints, but do not show signs
of the local hexagonal ordering. In contrast, $\mu_{5}$ serves as a direct
measure of %directly probes a
many-body structural coherence.

%\begin{figure}[htpb]
\begin{figure}[t]
\centering
\includegraphics[angle=-90, width=\columnwidth]{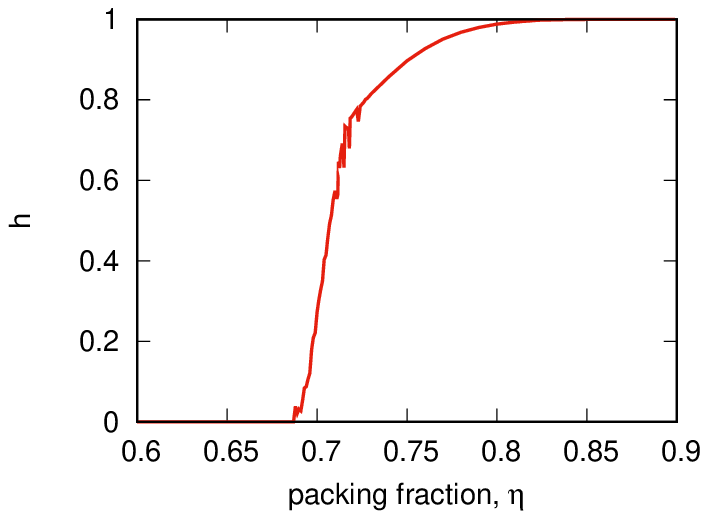}
\caption{The local scalar indicator of hexagonal ordering, $h$, as a
function of packing fraction $\protect\eta$.}
\label{fig:8}
\end{figure}

The non-monotonic behavior of $\mu _{5}$ has a clear geometric origin. A
finite five-disk intersection area requires a local deviation from the
perfect hexagonal symmetry, since in an ideal triangular lattice it reduces
to a single point. The initial increase of $\mu _{5}$ with density reflects
the growing probability of forming geometrically frustrated local
structures, whereas its subsequent decrease reflects the progressive
elimination of distortions of local hexagonal patterns as the system
approaches crystalline order. A perfect, tightly packed triangular lattice
restricts five $\sigma $-circles to intersect at a single dimensionless
point; a finite intersection area strictly requires a violation of perfect
hexagonal symmetry (as detailed in Appendix~\ref{app:lattice} and
illustrated in Fig.~\ref{fig:9}). Consequently, we introduce the
quantitative indicator $h(\eta )$ of local hexagonal order : 
\begin{equation}
h(\eta )=1-\frac{\mu _{5}(\eta )}{\mu _{5,\max }}\,.  \label{h}
\end{equation}%
The function $h(\eta )$ (see Fig.~\ref{fig:8}) varies smoothly from zero at
the ordering onset to unity at close packing in line with the continuous
character of the hexatic-to-solid transformation. In this sense, $h(\eta )$
provides a purely geometric and parameter-free scalar measure of the local
hexagonal ordering.

The orientation order parameter $\Psi_{6}$ for a 2D crystal has been
introduced by Nelson and Halperin \cite{Psi6} as its spatial correlations
reflect the quasi-long range ordering associated with the hexatic phase. It
has also been adopted in the HD model for the hexatic phase and its
correlation behavior has been established in the HD simulations \cite{ref11}%
. In harmonic 2D solids "a liquid close to its freezing point has a local
structure similar to that of a solid" \cite{ref36}. In other words, the
strong force keeps atoms at the sites of the solid's lattice, while the {%
long-wavelength} fluctuations only slightly modulate its primitive cells.
But a hard disk system is a soft matter. Indeed, hard disks move
ballistically, and the local structure consisting of a disk and its six
nearest neighbors can vary considerably on the scale of the figure formed by
these seven atoms. Clearly, at densities above the onset of the hexatic
phase, $\eta \sim 0.7$, the above seven disk figures are close to hexagons,
and as the density increases, so does their quality. If the order in local
hexagons changes, then it is natural to quantify it and the quantity $h$,
Eq.~(\ref{h}), can be used as its scalar measure. 
%We emphasize that $h$ \textbf{should not be regarded as an} order parameter \textbf{in the thermodynamic sense,}
%We emphasize that the $h$ is not an order parameter as the order is encoded in the spatial two-point correlation function, which for $h$ is trivial and factorizes to the product $h(x)h(x^{\prime }).$

In general, the introduction of an order indicator is a nontrivial and
ambiguous problem. Particularly challenging is choosing an order indicator
that can adequately reflect the need to incorporate many-body effects (e.g.,
the choice of the order parameter $Q_{6}$ to identify the structural
precursors of phase transitions and glassy dynamics in \cite%
{russo2015assessing,leocmach2013importance}). Luckily, the parameter $h(\eta)
$ {naturally} emerges as a byproduct of our theory directly from the
underlying many-body geometrical correlations {encoded} in the free-volume, 
%\textbf{construction,}
%directly from the underlying many-body geometrical correlations in the free volume, 
offering local characterization of the hexagonal order complementary to the $%
\Psi_{6} $. As required, our microscopic theory provides the mean value of
the observable $\mu _{5}$, Fig.\ref{fig:3}. The identification of $\mu _{5}$
as a structural indicator is consistent with the modern idea that many-body
correlations play a central role in determining the behavior of dense
liquids. In particular, it aligns with Tanaka's perspective that
medium-range crystalline order and locally favored structures govern the
emergence of slow dynamics and phase transitions in glass-forming systems.
Within the present framework, such many-body effects arise naturally from
the geometry of the free volume without the need to introduce additional
structural fields or phenomenological parameters.

\section{Conclusions}

We have developed a geometrical formulation of the statistical mechanics of
a 2D monodisperse hard-disk system in which the configurational PF is
determined by the free volume. The crucial result enabling this is the exact
analytical formulae obtained for the free volume as a function of disk
coordinates. This {establishes} a direct analytical {route} from the disks'
configurations to the free volume, cavity, and other geometrical objects so
that both insertions and intermediate constructions are of no need to
compute them. By expressing both free volume fractions, i.e., the extensive
cavity and intensive private cell in terms of the multi-disk intersection
areas, we reduce the evaluation of the configurational integral to a finite
set of constituting geometric quantities. Namely, because intersections of
more than five disks are forbidden by the hard-core constraint, the PF,
entropy, equation of state, and other thermodynamic quantities can be
expressed in terms of only four density-dependent functions, $\mu _{k}(\eta)$
with $k=2,3,4,5$. This establishes an 
%exact finite \textbf{analytical reduction}
exact finite closure of the many-body problem for the hard-disk model at the
level of five-body correlations. 
%An essential feature of the present formulation is that the configurational partition function is governed by the global excluded-volume geometry of the entire system rather than by local particle environments. This global geometrical perspective naturally leads to the decomposition of the free volume into the extensive cavity and intensive private cell and provides the basis for the analytical reduction developed here.
%or
%A key conceptual result is that the configurational partition function is determined by the global excluded-volume geometry of the system, from which the free volume, cavity, and private cell emerge naturally.

%\textbf{A central conceptual result of the present work is that the configurational partition function is determined by the topology of the global excluded-volume geometry rather than by local particle environments. This observation enables its exact analytical reduction to four constitutive geometrical measures, $\mu_2,\mu_3,\mu_4,\mu_5$, which completely determine the free volume, entropy, equation of state, and related thermodynamic properties.}

Crucially, the present geometric framework naturally incorporates many-body
correlations and demonstrates how they emerge directly in the free-volume
description beyond the scope of classical two-body approaches. Within this
framework, we obtained a factorized representation of the PF and identified
its two limiting asymptotic regimes %identified its two limiting regimes 
governed by the topology of the free volume: a low-density regime dominated
by connected cavities and a high-density regime characterized by
disconnected, particle-centered cells. Using the intersection-based
geometric measures analytically calculated from a representative equilibrium
particle configuration, the resulting theory reproduces the established
equation of state over nearly the entire density range up to close packing.
In the intermediate regime, we identified a mixed-fluid region in which
localized caged nearly hexagonal-shaped configurations in the highly
constrained environments contribute an additional configurational entropy
providing a natural interpretation of a defect-mediated structural
heterogeneity preceding the hexagonal ordering.

More broadly, the present results demonstrate that the complexity of
hard-core systems does not necessarily require an infinite hierarchy of
correlation functions. Instead, this complexity can be captured by a finite
set of geometrical measures that encode many-body correlations in a compact
and physically transparent form. This perspective provides a 
%\textbf{complementary} 
%non-perturbative \textbf{geometrical formulation}
non-perturbative alternative to traditional approaches based on virial
expansions or approximate closures, and 
%suggests a general route toward formulating the statistical mechanics of dense systems in closed form. 
suggests a general route toward formulating the statistical mechanics of
dense hard-core systems in closed form. The same perspective suggests that {%
other} hard-core systems may admit similar finite closure determined by the
geometry of exclusion constraints.

%The same perspective suggests that hard-core systems more generally may admit similar finite closures with the order determined by the geometry of exclusion constraints.

By analytically {identifying} %resolving 
a mixed-fluid region governed by competing local orderings, our model
provides a theoretical framework for two-order-parameter scenarios of liquid
vitrification and frustration. Furthermore, the five-disk intersection area
serves as a novel %highly sensitive 
many-body structural metric for incipient crystallization. Because free
volume is a universal geometrical construct, the exact mathematical
principles and PF factorization derived here are {extendable} to the 3D
hard-sphere model. %fully transferable to the 3D hard sphere model. 
This provides a general pathway and analytical basis to address the
geometric and hard-core aspects in highly confined and densely packed soft
matter systems, frustration-driven glass transitions, and complex
self-assemblies.

We outline a few important directions to address which follow from this
work. First, to smoothly describe the crossover between the lower density
and higher density asymptotic limits which implies assessing the mechanism
of cavities' connectivity breaking and forming cages of private cells.
Second, to establish a self-contained thermodynamic relation between the
defect fraction in the mixed-fluid regime and density. 
%and the intersection-based geometric measures. 
Third, derivation of the functions $\mu _{k}(\eta )$ 
%\textbf{constitutes} a well-posed, \textbf{albeit challenging}, mathematical problem
is a well-posed {albeit challenging} mathematical problem 
%(but probably very difficult) new mathematical problem 
which in principle can be addressed without resorting to disk coordinates.
Solving it would make the theory fully analytical. Finally, the geometrical
framework developed here can be extended to the 3D hard-sphere model where 
%a similar but higher-order 
%\textbf{finite analytical reduction} is expected.
a similar but higher-order finite closure is expected. More generally, the
intersection-based approach offers a promising route for describing the
thermodynamics and structural organization of dense hard-core systems,
including those exhibiting frustration, confinement, or glassy dynamics.

\section*{Acknowledgments}

This article is dedicated to Hajime Tanaka, honoring his extraordinary and
lasting impact on the physics of soft and liquid matter. His unifying
concepts, particularly the two-state models of liquids and the critical role
of medium-range crystalline order \cite{tanaka2022roles}, provided
foundational inspiration for formulating the geometric many-body approaches
detailed in this work. A.T. also gratefully acknowledge Hajime Tanaka's
interest in this work and valuable discussions during the Andalo 2026
International Workshop on Complex Systems.

VMP is grateful to the Center for Theoretical Physics PAS for hospitality.
This research is part of the project No. 2022/45/P/ST3/04237 co-funded by
the National Science Centre and the European Union Framework Programme for
Research and Innovation Horizon 2020 under the Marie Sk\l odowska-Curie
grant agreement No. 945339. T.B. was supported by the National Research
Foundation of Ukraine Project No. 2023.05/0019. A.T. thanks Barbara
Hribar-Lee for the invitation and gratefully acknowledges the hospitality of
the Faculty of Chemistry and Chemical Technology of the University of
Ljubljana and financial support through the MSCA4Ukraine project, funded by
the European Union (grant agreement ID: 1039539).

\appendix

\section{Analytical Formulas for Multiple Disk Intersections}

\label{app:intersections}

%To compute the intersection measures $\mu_k$ entering the free-volume expressions analytically from the particle coordinates, we adapt Kratky’s geometrical formulas~\cite{ref38,ref39}.
To analytically compute the free volume from particle coordinates, we adapt
Kratky's geometrical formulas~\cite{ref38,ref39} to the specific constraints
of the hard-disk model. Let the hard cores have a radius of $\sigma/2$, and
the effective $\sigma$-circles have a radius of $\sigma$. The quantities $I$
introduced below are the elementary intersection areas whose sums define the
measures $\mu_k$. %used in the main text.

\textbf{Intersection of Two $\sigma$-Circles ($I_{mn}$):} The intersection
area $I_{nm}$ of two $\sigma$-circles $m$ and $n$ separated by distance $%
r_{nm}$ is: 
\begin{equation}
I_{nm}=2\sigma ^{2}\left[ \cos ^{-1}\left( \frac{r_{nm}}{2\sigma }\right) -%
\frac{r_{nm}}{2\sigma }\sqrt{1-\left( \frac{r_{nm}}{2\sigma }\right) ^{2}}%
\right] ,  \tag{A1}
\end{equation}%
for $0\leq r_{nm}\leq 2\sigma $, and zero otherwise.

\textbf{Intersection of Three $\sigma$-Circles ($I_{mnl}$):} For three
mutually overlapping $\sigma$-circles, with all pair separations $r\leq
2\sigma $, %For three mutually overlapping disks ($r\leq 2\sigma $), 
we evaluate the circumradius $R_{mnl}$ of the triangle $\Delta (nml)$ formed
by their centers. If $R_{mnl}<\sigma $: 
\begin{equation}
I_{mnl}=\frac{1}{2}(I_{mn}+I_{ml}+I_{nl}-\pi \sigma ^{2})+\frac{%
r_{mn}r_{ml}r_{nl}}{4R_{mnl}}.  \tag{A2}
\end{equation}
If $R_{mnl}\geq \sigma $, the common intersection depends on the position of
the circumcenter $C_{mnl}$. If $C_{mnl}$ lies inside $\Delta (nml)$, the
triple intersection vanishes, $I_{mnl}=0$. If $C_{mnl}$ lies outside, the
common area $I_{mnl}$ reduces to the pair-intersection area associated with
the longest side (e.g., if $mn$ is the longest, $I_{mnl}=I_{mn}$).

%If $R_{mnl}\geq \sigma $, the area depends on the position of circumcenter $C_{mnl}$. If $C_{mnl}$ is inside $\Delta (nml)$, $I_{mnl}=0$. If $C_{mnl}$ is outside, $I_{mnl}$ collapses to the pair intersection of the largest side (e.g., if $mn$ is the largest, $I_{mnl}=I_{mn}$).

\textbf{Intersection of Four $\sigma$-Circles ($I_{knlm}$):} Considering a
quadrangle made of four $\sigma$-circle centers $knlm$, let $r_{nm}$ and $%
r_{kl}$ be the diagonals with $r_{nm}>r_{kl}$. If the circumradia of all
four constituent triangles are below $\sigma$: 
\begin{equation}
I_{knlm}=I_{nkl}+I_{mkl}-I_{kl}.  \tag{A3}
\end{equation}%
If any circumradius (e.g., $R_{knm}$) is $\geq \sigma $, then $%
I_{knlm}=I_{mnl}$ if the circumcenter $C_{knm}\notin \Delta (nmk)$, and $%
I_{knlm}=0$ if $C_{knm}\in \Delta (nmk)$ or is on its perimeter.

\textbf{Intersection of Five $\sigma$-Circles ($I_{knlm5}$):} For five $%
\sigma$-circles obtained by adding a fifth center to a valid four-center
configuration $knlm$ 
%For pentagons formed by adding a fifth disk $5$ to a valid quadrangle $knlm$ 
(where $r_{nm}>r_{kl}$), if all underlying triangle circumradia are bellow $%
\sigma$: 
\begin{equation}
I_{knlm5}=I_{mkl5}+I_{nkl5}-I_{kl5}.  \tag{A4}
\end{equation}
If any constituent triangle has circumradius equal or above $\sigma$ (e.g., $%
R_{nlm}\geq \sigma $), the intersection reduces to the four-disk case $%
I_{onm5}$ if $C_{nlm}\notin \Delta (nlm)$, and vanishes if $C_{nlm}\in
\Delta (nlm)$ or is on its perimeter.

\section{Vanishing of the Cavity and Intensive Cell in a Densely Packed
Triangular Lattice}

\label{app:lattice}

Here we demonstrate how to identify and compute the areas of all multiple
intersections of the $\sigma$-circles associated with a representative disk
in the close-packed triangular lattice. 
%As all disks in the lattice are similar, the averaging over the configuration is not necessary.

\begin{figure}[htpb]
%\begin{figure}[b]
\centering
\includegraphics[width=0.45\columnwidth]{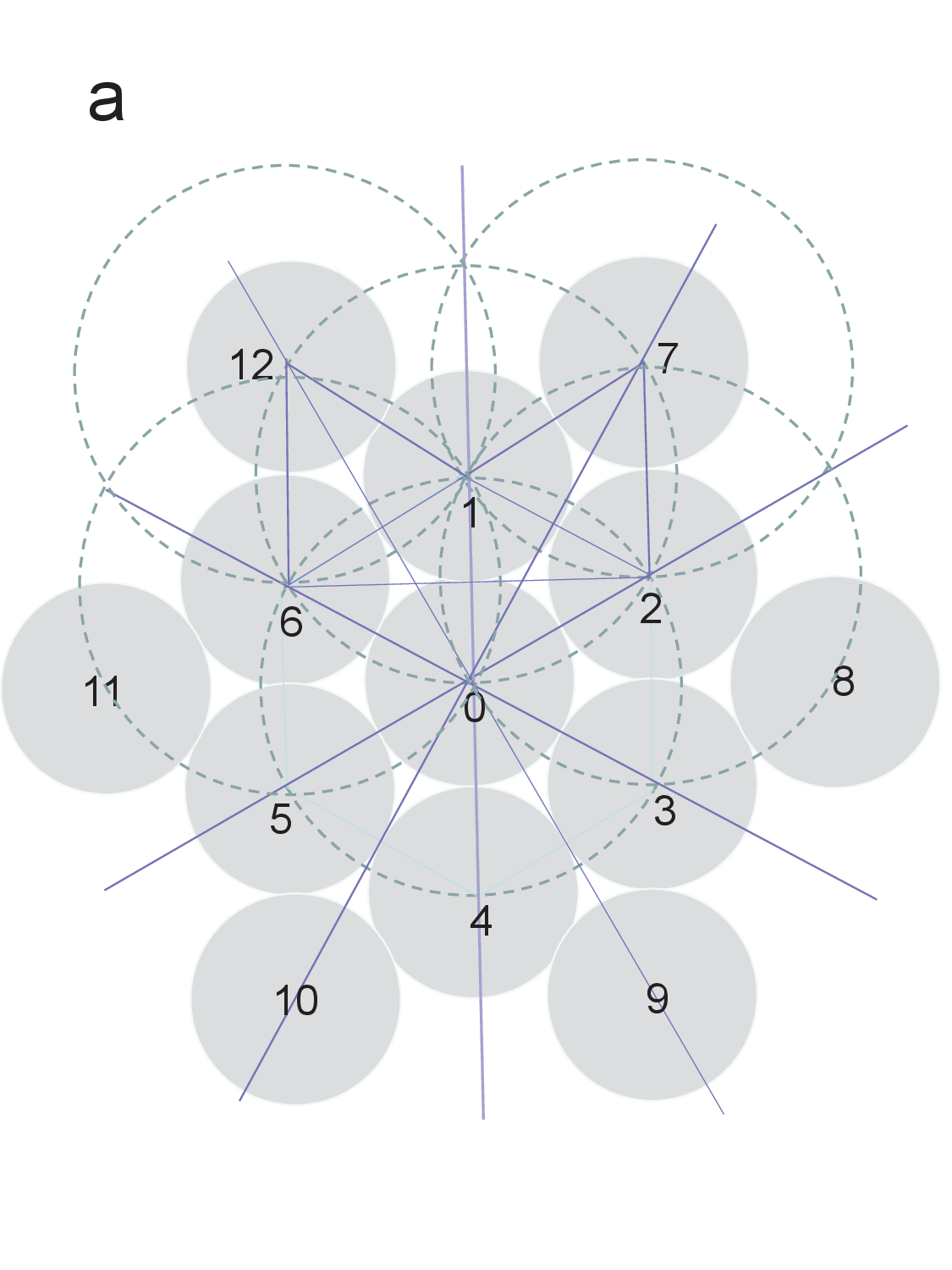} %
\includegraphics[width=0.45\columnwidth]{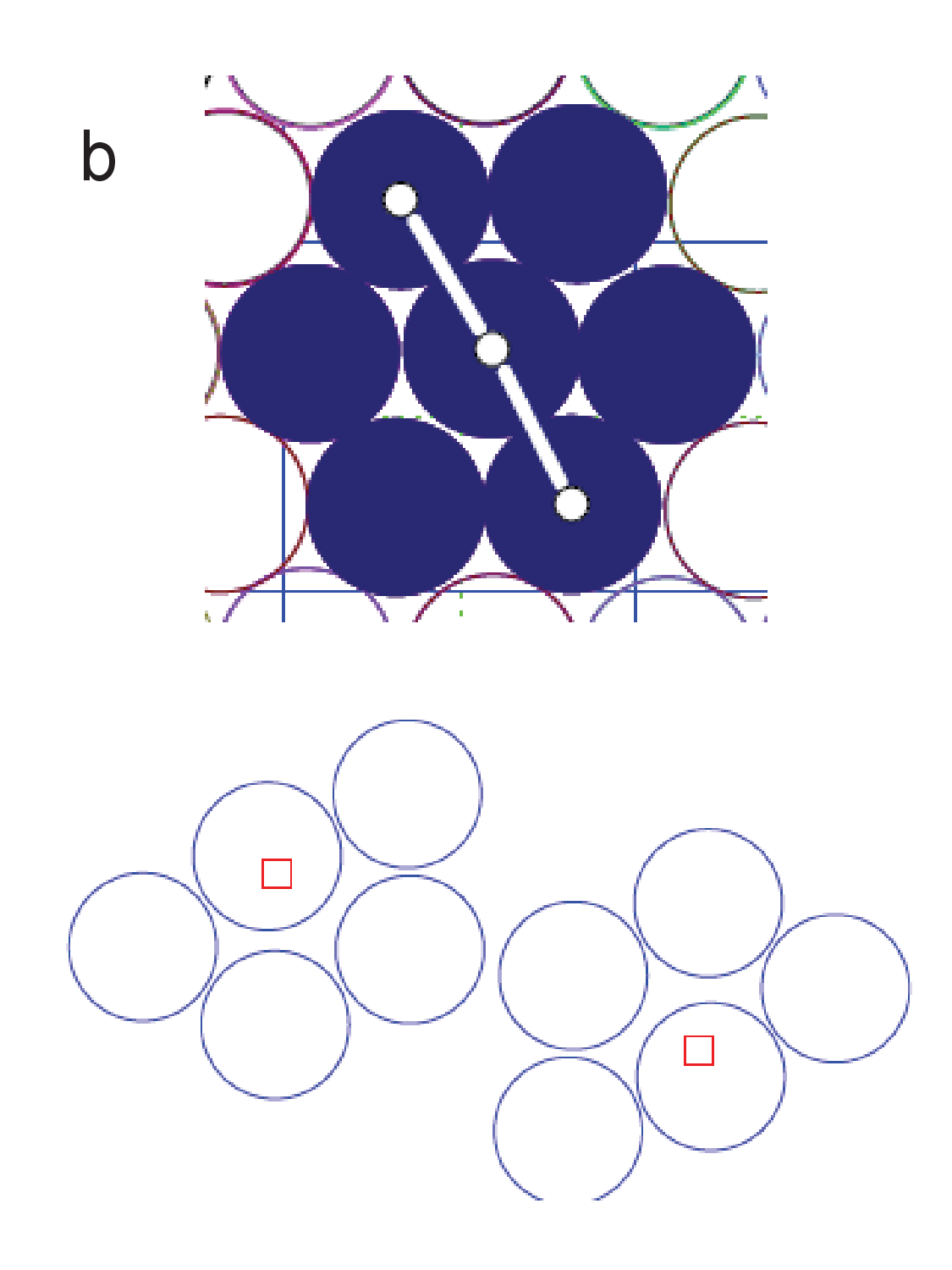}
\caption{Part a: A fragment of a triangular densely packed hard disk
lattice. The disks' cores are filled circles of the radius $\protect\sigma/2$%
. Five $\protect\sigma$-circles centered at 0, 2, 7, 12, and 6 are indicated
by dashes. The central disk 0 has next neighbors centered at 1, 2, 3, 4, 5,
and 6, and next-next neighbors centered at 7, 8, 9, 10, 11, and 12. The $%
\protect\sigma$-circles of all disks in the lattice, which are not shown, do
not overlap with the $\protect\sigma$-circle centered at 0. The common
intersection area of five $\protect\sigma$-circles, e.g., 02165, 02834, is a
point. Part b: Five $\protect\sigma$-circles common intersection area can be
finite only if the disks' hexagonal order is violated. In the upper panel,
the white line is not straight as it would be in a perfect triangular
lattice, and the common intersection area of the $\protect\sigma$-circles of
the five disks on the left of the line is nonzero. In the lower panel, the
common intersection areas of the $\protect\sigma$-circles of the two sets of
five disks are shown by the squares.}
\label{fig:9}
\end{figure}

A fragment of this lattice is shown in Fig.~\ref{fig:9}a. The hard cores of
the disks, which are in contact with each other, have a radius of $\sigma /2$%
. The attached concentric $\sigma$-circles of radius $\sigma$ are indicated
by dashed lines (shown here for five disks to prevent visual clutter). The
central disk (labeled 0) is shown along with its six nearest neighbors
(1--6) and its six next-nearest neighbors (7--12). The {center-to-center
distance to the next-nearest neighbors is less than $2\sigma$,} meaning that
their $\sigma$-circles overlap with the central $\sigma$-circle. No other
disks in the lattice have $\sigma$-circles that overlap with disk 0 .

As illustrated in Fig.~\ref{fig:9}, in a perfect lattice, five $\sigma$%
-circles intersect exactly at a single %dimensionless 
point; a finite common intersection area can only emerge if the perfect
hexagonal order is violated.

To compute the total intersection areas for the central disk, we find and
list all different pairs, triples, and quadruples of $\sigma$-circles that
include disk 0, compute their areas $S$, and sum them.

\begin{itemize}
\item \textbf{Pairs ($S_2$):} There are six pairs similar to (0,1) and six
pairs similar to (0,7). The area $S_{01}$ is bounded by circle 0 from above
and circle 1 from below, giving $S_{01} = 2(\pi/3 - \sqrt{3}/4)\sigma^2$.
The area $S_{07}$ is the lobe with vertices 1 and 2, giving $S_{07} =
2(\pi/6 - \sqrt{3}/4)\sigma^2$. The total contribution is $S_2 = 6S_{01} +
6S_{07}$.

\item \textbf{Triples ($S_3$):} The total area is $S_3 = 6S_{102} +
12S_{017} + 6S_{602} = 6S_{102} + 18S_{017}$. The area $S_{102}$ is the
curvilinear triangle, $S_{102} = [\sqrt{3}/4 + 3(\pi/6 - \sqrt{3}%
/4)]\sigma^2 $. The area $S_{017}$ equals $S_{07}$.

\item \textbf{Quadruples ($S_{4}$):} The total area is $%
S_{4}=6S_{0216}+6S_{0172}=12S_{0216}$. The area $S_{0216}$ is the lobe with
vertices 0 and 1, which equals $S_{07}$.
\end{itemize}

%Using these geometrical contributions, we evaluate the close-packing value of the intensive cell area
Using these specific geometric areas, we compute the close-packing value of
the intensive cell volume $c_{\mathrm{cp}}=\pi \sigma ^{2}-\mu _{2}+\mu
_{3}-\mu _{4}+\mu _{5}$. As $\mu_{5}$ is a %dimensionless 
single point in the perfect lattice, it evaluates to zero. 
%Because $\mu_{5}$ reduces to a single point in the perfect lattice, its area contribution is zero.
Substituting the sums derived above one finds: 
\begin{equation}
c_{\mathrm{cp}}/\sigma ^{2}=\pi -\left[ 6(\pi /6-\sqrt{3}/4)+2\pi -3\sqrt{3}%
/2\right] =\pi -\pi =0.  \tag{B1}
\end{equation}

Similarly, to evaluate the extensive cavity, we divide $C_{\mathrm{cp}}$ by $%
N\sigma ^{2}$ to address the size-independent quantity. Using $%
V/(N\sigma^{2})=\pi /(4\eta _{\mathrm{cp}})$, where the close-packing
fraction is $\eta_{\mathrm{cp}}=\pi/(2\sqrt{3})$, we find: 
\begin{equation}
C_{\mathrm{cp}}/(N\sigma ^{2})=\pi /(4\eta _{\mathrm{cp}})-\pi +\pi -\sqrt{3}%
/2=0.  \tag{B2}
\end{equation}

This confirms that not only does the total free volume vanish at close
packing, but both intensive cell and extensive cavity vanish separately,
proving their functional independence. 
%This \textbf{demonstrates that, at close packing,} not only does the global free volume vanish, but \textbf{its two constituent contributions -- the intensive private cell and the extensive cavity -- also vanish independently. This confirms that they represent complementary geometrical contributions to the same global free volume rather than alternative definitions of it.}

\section{Partition Function in Terms of the Free Volume}

\label{app:partition}

\subsection{Factorization of the General PF}

Here we first show that the integration domains in the configurational
partition function $Z$, Eq.~(\ref{ZNN}), are precisely the free volumes, and
then {demonstrate} that $Z$ factorizes into their product. 
%Here we first show that the integration domains in the PF $Z$ (\ref{ZNN}) are the free volumes and then prove that $Z$ factorizes into their product.
Consider the $k$-th integral $I_{k}$ in the general expression for the PF: 
\begin{equation}
I_{k}=\int_{V}dx_{k}\exp \left( -\sum_{i=1}^{k-1}U_{ik}\right) .  \tag{C1}
\end{equation}%
The hard-core potential $U_{ik}$ dictates that the center $x_{k}$ of a
circle $\sigma _{k}$ cannot enter any $\sigma _{i}$ with $i<k$. The circles $%
\sigma_{i}$ can be defined as sets of points $x_{k}$: $\sigma
_{i}=\{x_{k}:|x_{i}-x_{k}|\leq \sigma \}$. We define the indicator function $%
\tau _{i}$ of this set as $\tau _{i}=1$ if $x_{k}\in \sigma _{i}$ and $\tau
_{i}=0$ otherwise. The product of $n$ indicators for different $\sigma _{i}$
is the indicator of their intersection set. Noting that $\exp
(-U_{ik})=1-\tau _{i}$, the exponential in $I_{k}$ can be expanded into
products of such terms: 
\begin{align}
\exp \left( -\sum_{i=1}^{k-1}U_{ik}\right) & =\prod_{i=1}^{k-1}(1-\tau _{i})
\tag{C2} \\
& =1-\sum_{i=1}^{k-1}\tau _{i}+\sum_{i>j}^{k-1}\tau _{i}\tau
_{j}-\sum_{i>j>l}^{k-1}\tau _{i}\tau _{j}\tau _{l}+\dots  \notag
\end{align}%
As a maximum of five $\sigma $-circles can intersect without their hard
cores overlapping, all products of more than five $\tau $'s vanish.
Integrating both sides\ of (C2) over $V$ regarding the definitions (\ref%
{Vaverage}) and (\ref{cNn}) directly yields the free volume: 
\begin{equation}
I_{k}=V_{k}\{x_{k-1}\}_{N}\,.  \tag{C3}
\end{equation}
%
%\textbf{Equation (C3) establishes the exact analytical connection between particle coordinates and the global free volume, which forms the basis of the subsequent factorization of the configurational partition function.}
This is the conditional free volume of the $k$-th disk in a maze made by a
configuration $\{x_{k-1}\}_{N}$ of ($k-1$) disks, which is subject to the
accommodation condition that it leaves space for the rest ($N-k+1$) disks no
matter what their configuration is. To establish factorization, we introduce
the integral operator $Z_{N-1,N}$: 
\begin{equation}
Z_{N-1,N}(...)={\int_{V}}^{\prime }dx_{1}\dots dx_{N-1}\exp \left(
-\sum_{i>j}^{N-1}U_{iN}\right) (...)\,,  \tag{C4}
\end{equation}
where the prime denotes integration over all allowed configurations that
leave space for disk $N$. The integral $Z$ can then be presented as the
action of $Z_{N-1,N}$ on $V_{N}$: 
\begin{equation}
Z\propto Z_{N-1,N}\{x_{N-1}\}_{N}(V_{N}\{x_{N-1}\}_{N})\,.  \tag{C5}
\end{equation}
%
%For Sinai hard-disk systems, equilibrium dynamics possesses the property of K-mixing. Consequently, extensive observables converge to their equilibrium values independently of the initial configuration.
The hard-disk system has the statistical property of $K$-mixing \cite{ref32}%
. In $K$ systems, in thermodynamic equilibrium an extensive observable $O$
converges to its limiting value $\langle O\rangle $ (and not only to its
average value as in ergodic systems without mixing) \cite{KSF}. Therefore,
the cavity $C\{x_{N-1}\}_{N}$ converges to its limiting equilibrium
expectation value $\left\langle C_{N+1}\right\rangle _{N}$ which is the same
for all configurations $\{x_{N-1}\}_{N}$. Then the cavity contribution to
the above integral over all configurations reduces to the product 
%\textbf{Then the cavity contribution therefore reduces, in the thermodynamic limit, to the product}
\begin{equation}
Z_{N-1,N}\{x_{N-1}\}_{N}(C\{x_{N-1}\}_{N})=\left\langle C_{N+1}\right\rangle
_{N}\times Z_{N-1,N}\,.  \tag{C6}
\end{equation}
%
%\textbf{Thus both the extensive cavity and the intensive private-cell contribution enter the factorized partition function through their equilibrium averages.}
Now consider $c_{N}\{x_{N-1}\}_{N}$. Its contribution to $V_{N}$ depends on
the coordinate $x_{n}$ and specific $\{x_{N-1}\}_{N}$, but the total sum
over all $\{x_{N-1}\}_{N}$ of course does not. To find this constant value $%
\langle c_{N}\rangle $ we notice that the result will be same if $x_{n}$ is
swapped with any $x_{i}$, $i=1,2,...,N-1$. It follows that $\langle
c_{N}\rangle =(1/N)\sum c_{n,N},$ the expected result which is formula (\ref%
{cN}). Combining this with (C6), the right hand side of (C5) becomes $%
\left\langle V_{N}\right\rangle _{N}\times Z_{N-1,N}$ where $\left\langle
V_{N}\right\rangle _{N}=\left\langle C_{N+1}\right\rangle _{N}+\left\langle
c_{N}\right\rangle$. %
%Repeating the same procedure for $k=N-1, N-2, \ldots$
Continuing along this line for $k=N-1, N-2, \ldots$ one obtains $Z$ (\ref%
{ZNN}) in the form of a product of the average free volumes.

\subsection{Relation between Disks' Permutations, Lattice Model, and
Logarithmic Pressure Divergence}

It is important to realize that $Z_{G}$ (\ref{ZG}) and the general PF in the
form (\ref{ZNN}), if treated formally as the integral over the whole volume $%
V$, 
%\textbf{cannot reproduce the correct high-density (actually $\eta$ above $0.53$) pressure behavior.}
cannot give the correct pressure for $\eta $ above $0.72$ (actually, above $%
0.53$). The key to seeing this is in observing that, if treated formally,
expression (\ref{ZNN}) is a continuous generalization of a factorial and
intimately related to permutations and following lattice model: $N$
particles on $J$ lattice sites, no more than one on each site of size $V/J$,
where $V$ is the total volume. The exact PF of this model is 
%This follows from the observation that Eq.~(\ref{ZNN}), when interpreted as an integral over the entire volume, is a continuous analogue of a factorial and is closely related to the following lattice model:
%
\begin{equation}
Z_{lat}=(V/J)^{N}J!/(J-N)!N!  \tag{C7}
\end{equation}
and the resulting pressure is 
\begin{equation}
P_{J,N}=-(N/V)ln[(V/N)(1-N/J)].  \tag{C8}
\end{equation}
At the close packing, $N\rightarrow J$, this pressure diverges as $\ln($free
volume$)$, the same way as the pressure of the gas approximation, Fig.6,
whereas the correct pressure, both $P_{exp}$ and $P_{L}$, diverges as $1/($%
free volume$)$. 
%\textbf{The lattice model illustrates that a logarithmic divergence is the general consequence.}
The lattice model shows that a logarithmic divergence is the general
corollary of the integration over the progressively decreasing areas,
starting from the full volume in $Z$ (\ref{ZNN}) and $Z_{G}$ (\ref{ZG}), and
over all $J$ sites in the lattice model. 
%\textbf{The failure therefore originates not from the free-volume representation itself, but from extending the full configurational integration beyond the density range where unrestricted particle permutations remain dynamically relevant.}
%\textbf{This behavior pertains to the "gas-like" (or G) density region,}
This is pertinent to the "gas" densities, when the cavity consists of large
connected pieces which are accessible for disks and where they can exchange
their positions during relaxation time $t_{\mathrm{rel}}$ (see next
subsection).

%\textbf{The lattice analogy discussed here serves only to illustrate the role of unrestricted particle permutations; it is not intended as a microscopic model of the hard-disk fluid. The essential point is that unrestricted particle permutations imply integration over a connected global cavity. The lattice model provides a simple example showing that such unrestricted integration necessarily leads to a logarithmic divergence of the pressure near close packing.}

\subsection{The Role of Relaxation Time}

%\textbf{To clarify the role of particle permutations in the configurational partition function, we introduce the relaxation time $t_{\mathrm{\rm rel}}$.}
Considering status of disks' permutation in the PF integral we resorted to
the relaxation time $t_{\mathrm{rel}}$ which we now want to clarify. To see
if disks exchange their positions in the equilibrium state at a given
density we need to see a configuration at different times. This is time on
order on the relaxation time $t_{\mathrm{rel}}$ of our system at a given
density. As explained in Appendix C.2, the dynamic quantities of a hard disk
system, possessing the $K$ mixing property, relax to their equilibrium
values over the time $\sim t_{\mathrm{rel}}$: over each $t_{\mathrm{rel}}$
time the system moves from one equilibrium configuration to another
statistically independent one %equilibrium configuration 
(for shorter time these would correlate) and each results in the same mean
values. Thus, one can assess the probability of coordinate swapping by
estimating the number of the swapping disk pairs after time $\sim t_{\mathrm{%
rel}}$. For instance, if after $t_{\mathrm{rel}}$ we found a 100 such pairs
in a configuration of 10$^{6}$ disks, then the probability is $10^{-4}$ and
averaging over time we will find a fluctuation from the fully caged state
with this weight $10^{-4}$. However, over the time $\sim t_{\mathrm{rel}}$
the system does not sample all possible states but only those that
correspond to the single actual trajectory with otherwise arbitrary initial
configuration. It is in order to explain this in terms of the PF which is
the formal sum over "all states". 
%\textbf{The PF itself is independent of dynamics; the relaxation time is introduced only to determine which configurations contribute with non-negligible statistical weight in the corresponding asymptotic regime.}
The PF (\ref{ZNN}) samples states in arbitrary sequence, i.e., randomly from
trajectories with all possible different initial configurations. 
%\textbf{The relaxation time therefore does not modify the definition of the configurational partition function; rather, it provides the physical criterion for identifying the asymptotically relevant configurational states in different density regimes.}
Clearly, such sampling includes all disk permutations which then must be
accounted for by the factor $1/N!$. At the same time, as all equilibrium
states result in the same averages, depending on the convenience one can
consider the PF either sampling all states or the state of a single
equilibrium trajectory. When a system is dense (the L approximation), then a
single trajectory gives the full information: as disks' position swapping is
dynamically suppressed, the $N!$ permutations produce $N!$ statistically
identical (up to disk vibrations in their cages) states resulting in the PF $%
Z_{L}$ (\ref{ZL}) with the prefactor $N!/N!=1$. At a low-density regime (the
G approximation), the situation is very different: because of many swapping
during $t_{\mathrm{rel}}$, the state at $t=0$ differs from the state at $nt_{%
\mathrm{rel}}.$ Then to incorporate distinguishable states it is simpler to
formally %\textbf{sampling all admissible configurational}
sample all possible states including indistinguishable ones and then divide
the sum by $N!$\,.

We see that it is sometimes convenient to address a PF as not the sum over 
\emph{a priori} all configurations, but that over typical states which are
present in the real system with a sufficiently nonzero probability. This
approach might seem unusual, but actually it has been explicitly used and
considered natural. Here is an example. Consider a system of $N$ particles
in a volume $V.$ The PF is the volume integral over the particles'
coordinates. {Correct.} Now we add that this system is a crystal and these
particles stay at its sites. 
%\textbf{For a crystal, one naturally restricts the configurational integration to small-amplitude vibrations about lattice sites rather than integrating over the entire volume. The underlying reason is the same as in the dense hard-disk system: particle exchanges occur on time scales much longer than the characteristic relaxation time and therefore carry negligible statistical weight.}
Would anybody really think of integrating particles' coordinates over the
entire $V?$ Of course not: instead, it is natural to consider 
%the small amplitude  of the particle 
small amplitude particles' vibrations at the lattice sites. The reason for
this restriction of the integration domain is exactly the same as in our
dense HD system: swapping of sites is possible, but it is so a rare event
that during the characteristic crystal time one cannot see it. Thus, the
configurations in the PF should be given weights corresponding to the system
dynamics. To summarize, at high densities, the system produces states that
differ only by positions of disks in their cages which is accounted for by
the mean free volume per disk.

For a dense system, it is also instructive to understand the relation
between the time averaging over a single trajectory and averaging over all
states in the PF (\ref{ZNN}). Given the %a particular 
configuration $\{x_{k-1}\}$, the general integration domain $V_{k}\{x_{k-1}\}
$ is a union of more than one connected area. At the same time, averaging
over a single trajectory, the integral over $x_{k}$ is restricted to just
one of them, say $v_{k}$. {In this case, the average $\left\langle
v_{k}\right\rangle $ is that over time dependent shape of the cage which
contains $x_{k}$}: sometimes this cage has a finite cavity, sometimes only
private cell, so that $\left\langle v_{k}\right\rangle $ which enters the PF 
$Z_{L}$ (\ref{ZL}) is the sum $\left\langle \overline{C}\right\rangle
+\left\langle c\right\rangle$ where $\left\langle \overline{C}\right\rangle$
is the time mean cavity per disk. Alternatively, averaging in terms of the 
%partition function 
PF (\ref{ZNN}) is the sum over different states and, in particular, $%
\left\langle v_{k}\right\rangle$ %\textbf{is obtained} 
obtains by summing over all connected areas in $V_{k}\{x_{k-1}\}$. Some of
them have nonzero cavities, others have only private cells, but the space
average is also $\left\langle \overline{C}\right\rangle +\left\langle
c\right\rangle$. In simple words, this is known as "time average is equal to
phase space average", which is the essence of the Birkhoff-Khinchin theorem 
\cite{KSF}. In our PF $Z_{L}$ (\ref{ZL}) we do not neglect mean cavity $%
\left\langle\overline{C}\right\rangle$, but our numerical results show that
the actual cavity is negligible compared to $\left\langle{c}\right\rangle$. 

\section{Simulation details}

\label{app:simulations} The disk coordinates required to evaluate the
intersection measures $\mu_2, \ldots, \mu_5$ were obtained from Monte Carlo
simulations of the two-dimensional hard-disk system. 
%The disk coordinates that are necessary to compute the intersection area functions $\mu _{2}$ through $\mu _{5}$ were obtained using Monte Carlo computer simulations of two-dimensional hard disk system. 
The number of particles was fixed at $N=900=30\times30$ throughout the
simulations. The packing fraction $\eta={\pi N\sigma^2}/{(4V)}$ was varied
by changing the simulation cell area $V=L_x\times L_y$. 
%Thus, the density, i.e. packing fraction $\eta={\pi N\sigma^2}/{(4V)}$, was varied by changing the area $V=L_x\times L_y$ of the simulation cell. 
We used a rectangular cell with fixed aspect ratio $L_x:L_y = 1:\sqrt{3}/2$
and periodic boundary conditions in both directions. This geometry is
commensurate with a $30\times30$ triangular lattice, although it does not
eliminate finite-size or box-shape effects.

Equilibrium configurations were generated in the canonical ($NVT$) ensemble
using the conventional single-particle Metropolis algorithm for hard disks.
In one trial move, a disk was selected randomly and given a random
displacement drawn uniformly from a square centered on the disk. The move
was accepted if the displaced disk did not overlap any other disk and was
rejected otherwise. One Monte Carlo cycle is defined as $10^3$ attempted
single-particle moves. The maximum displacement was adjusted 
%[state how and how often] 
to maintain an acceptance probability of approximately $30\%$ to $50\%$.
Initial simulations were performed at the eight reference packing fractions $%
\eta=0.2,0.3,\ldots,0.9$. At each reference density, the initial $30\times30$
configuration was a perfect triangular arrangement scaled to the
corresponding cell dimensions. After equilibrating these reference states,
simulations were performed over the intervening density intervals with
various density step $\Delta\eta$ depending on the density range $\Delta\eta$
as specified below. Each intervening state was initialized from the
equilibrated configuration at the nearest higher packing fraction by
increasing $L_x$ and $L_y$ uniformly while preserving the aspect ratio.

In total 229 thermodynamic states in the density range $0.100\leq \eta \leq
0.900$ were investigated with varied density steps depending on the density
interval: $\Delta \eta =$ 0.01 ($0.100\leq \eta \leq 0.670$), $\Delta \eta =$%
0.001 ($0.670<\eta \leq 0.730$), $\Delta \eta =$ 0.002 ($0.730<\eta \leq
0.860$) and $\Delta \eta =$ 0.001 ($0.860<\eta \leq 0.900$). At each
density, the system was equilibrated. The equilibration period lasted from $%
5\times 10^{6}$ to $10\times 10^{6}$ (depending on density range) Monte
Carlo cycles. The longest equilibration runs, performed at densities $\eta
>0.8$, contained up to $10^{8}$ cycles. Equilibration was assessed from the
stationarity of the first coordination number obtained by integrating the
radial distribution function during the equilibration run. 
%Equilibration was assessed by stationarity of the first coordination number obtained by numerical integration of the radial distribution function that was evaluated during the pre-equilibrium runs. 
After equilibration, the production averages were accumulated over $5\times
10^{6}$ Monte Carlo cycles. Configurations were stored after every 100
cycles, giving 50 000 stored configurations at each density. For each stored
configuration, the intersection measures $\mu _{2},\mu _{3},\mu _{4},\mu _{5}
$ were evaluated analytically and subsequently averaged over the ensemble.
The resulting averages of $\mu_2,\mu_3,\mu_4,\mu_5$ constitute the only
numerical input required by the analytical theory developed in the main text.

%Further, obtained intersection areas were used to calculate the theoretical pressures $P_{\mathrm{G}}$ and $P_{\mathrm{L}}$ as functions of the density by the entropy differentiation with respect to $V$: $P_{\mathrm{G,L}}=\partial S_{\mathrm{G,L}}/\partial V$ with $V=N/(4\eta/\pi )$ ($k_{B}T$ is set one).

%\newpage

\section*{References}

% new List of references


\begin{thebibliography}{99}
\bibitem{ref6} C. P. Royall, P. Charbonneau, M. Dijkstra, J. Russo, F.
Smallenburg, T. Speck, and C. Valeriani, Colloidal hard spheres: Triumphs,
challenges, and mysteries, Rev. Mod. Phys. \textbf{96}, 045003 (2024).

\bibitem{SpatialStat} Statistical Physics and Spatial Statistics, edited by
K.R. Mecke and D. Stoyan, Lecture Notes in Physics (Springer, Berlin, 2000).

\bibitem{ref14} B. Li, Y. Nishikama, P. H\"{o}llmer, L. Carillo, A. C.
Maggs, and W. Krauth, Hard-disk pressure computations - a historic
perspective, J. Chem. Phys. \textbf{157}, 234111 (2022).

\bibitem{ref14a} K. Hukushima and W. Krauth, Damage spreading and coupling
in spin glasses and hard spheres, Front. Phys. \textbf{12}, 1507250 (2024).

\bibitem{ref8} B. J. Alder and T. E. Wainwright, Phase transition in elastic
disks, Phys. Rev. \textbf{127}, 359 (1962).

\bibitem{ref11} E. Bernard and W. Krauth, Two-step melting in two
dimensions: first-order liquid-hexatic transition, Phys. Rev. Lett. \textbf{%
107}, 155704 (2011).

\bibitem{ref12} M. Engel, J. A. Anderson, S. C. Glotzer, M. Isobe, E. P.
Bernard, and W. Krauth, Hard-disk equation of state: First-order
liquid-hexatic transition in two dimensions with three simulation methods,
Phys. Rev. E \textbf{87}, 042134 (2013).

\bibitem{ref13} S. Karper and W. Krauth, Two-dimensional melting: From
liquid-hexatic coexistence to continuous transitions,

Phys. Rev. Lett. \textbf{114}, 035702 (2015).

\bibitem{ref9} B. J. Alder, W. G. Hoover, and D. A. Young, Studies in
molecular dynamics. V. High-density equation of state and entropy for hard
disks and spheres, J. Chem. Phys. \textbf{49}, 3688 (1968).

\bibitem{ref10} J. Kolafa and M. Rottner, Simulation-based equation of state
of the hard-disk fluid and prediction of higher-order virial coefficients,
Mol. Phys. \textbf{104}, 3435 (2006).

%\bibitem{Lebovitz} J. L. Lebowitz and O. Penrose, Convergence of virial
%expansions, J. Math. Phys. \textbf{5}, 841 (1964).

\bibitem{virial10} N. Clisby and B.M. McCoy, Ninth and tenth order virial
coefficients for hard spheres in D dimensions, J. Stat. Phys. \textbf{122},
15 (2006).

\bibitem{Feynman} R. P. Feynman, \textit{Statistical Mechanics} (Westview
Press, USA, 1998).

\bibitem{ref2} J.-L. Barrat and J.-P. Hansen, \textit{Basic Concepts for
Simple and Complex Liquids} (University Press, Cambridge, 2003).

\bibitem{ref3} \textit{Theory and Simulation of Hard-Sphere Fluids and
Related Systems}, edited by A. Mulero, Lecture Notes in Physics (Springer,
Berlin, 2008).

\bibitem{ref15} J. Hirschfelder, D. Stevenson, and H. Eyring, A theory of
liquid structure, J. Chem. Phys. \textbf{5}, 896 (1937).

\bibitem{ref17} O. K. Rice, On communal entropy and the theory of fusion, J.
Chem. Phys. \textbf{6}, 476 (1938).

\bibitem{Hoover72} W.G. Hoover, W.T. Ashurst, R. Grover, Exact dynamical
basis for a fluctuating cell model, J. Chem. Phys. \textbf{57}, 1259 (1972).

\bibitem{ref19} W. G. Hoover, N. E. Hoover, and K. Hanson, Exact hard-disk
free volumes, J. Chem. Phys. \textbf{70}, 1837 (1979).

\bibitem{ref20} R. Speedy, Accurate theory of the hard sphere fluid, J.
Chem. Soc., Faraday Trans. 2 \textbf{73}, 714 (1977)

\bibitem{Speedy80} R. Speedy, Statistical geometry of hard-sphere systems,
J. C. S., Faraday II, \textbf{76}, 693 (1980).

\bibitem{ref21} R. Speedy, Cavities and free volume in hard-disc and
hard-sphere systems, J. Chem. Soc., Faraday Trans. 2 \textbf{77}, 329 (1981).

\bibitem{Speedy91} R. Speedy and H. Reiss, Cavities in the hard sphere fluid
and crystal and the equation of state, Mol. Phys. \textbf{72}, 999 (1991).

\bibitem{Stillinger92} K. Sturgeon and F. Stillinger, Cavities in the
harddisk crystal: A Monte Carlo simulation study, J. Chem. Phys. \textbf{96}%
, 4651 (1992).

\bibitem{ref22} M. D. Rintoul and S. Torquato, Algorithm to compute void
statistics for random arrays of disks, Phys. Rev. E \textbf{52}, 2635 (1995).

\bibitem{ref24} S. Sastry, T. M. Truskett, P. G. Debenedetti, S. Torquato,
and F. H. Stillinger, Free volume in the hard sphere liquid, Mol. Phys. 
\textbf{95}, 289 (1998).

\bibitem{2014} J. Kim and J.S. Bong, Dynamics and spatial correlation of
voids in dense two dimensional colloids, J. Chem. Phys. \textbf{141}, 014502
(2014).

\bibitem{2015} M. Schindler and A.C. Maggs, Cavity averages for hard spheres
in the presence of polydispersity and incomplete data, Eur. Phys. J. E 
\textbf{38} , 97 (2015).

\bibitem{2016} V. Baranau and U. Tallarek, Chemical potential and entropy in
monodisperse and polydisperse hard-sphere fluids using Widom's particle
insertion method and a pore size distribution-based insertion probability,
J. Chem. Phys. \textbf{144}, 214503 (2016).

\bibitem{2023} R.L. Davidchack, A. Elmajdoub, and B. B. Laird, Analysis of
probability of inserting a hard spherical particle with small diameter in
hard-sphere fluid, J. Chem. Phys. \textbf{159}, 184503 (2023).

\bibitem{Isobe} D. Mugita, K. Souno, H. Koyama, T. Nakamura, and M. Isobe,
Simple and efficient methods for local structural analysis in polydisperse
hard disk systems, J. Chem. Phys. \textbf{160}, 174104 (2024).

\bibitem{ref31} The complex shape of the free volume has a fully convex
boundary which allowed Sinai to prove that the hard-disk system has the
property of K-mixing \cite{ref32,KSF} and thus does relax to the
(microcanonical) equilibrium state.

\bibitem{ref32} Y. G. Sinai, Dynamical systems with elastic reflections,
Russ. Math. Surv. \textbf{25}, 137 (1970).

\bibitem{KSF} I.P. Kornfeld, Y.G. Sinai and S.V. Fomin, \textit{Ergodic
Theory} (Springer, New York, 1981).

\bibitem{Widom} B. Widom, Some topics in the theory of fluids, J. Chem.
Phys. \textbf{39}, 2808 (1963).

\bibitem{ref4} G. Parisi and F. Zamponi, Mean field theory of hard spheres
glasses and jamming, Rev. Mod. Phys. \textbf{82}, 789 (2010).

\bibitem{ref5} P. Charbonneau, J. Kurchan, G. Parisi, P. Urbani, and F.
Zamponi, Glass and jamming transitions: from exact results to
finite-dimensional description, Annu. Rev. Condens. Matter Phys. \textbf{8},
1 (2017).

\bibitem{tanaka2022roles} H. Tanaka, Roles of liquid structural ordering in
glass transition, crystallization, and water's anomalies, J. Non-Cryst.
Solids: X \textbf{13}, 100076 (2022).

\bibitem{leocmach2013importance} M. Leocmach, J. Russo, and H. Tanaka,
Importance of many-body correlations in glass transition: An example from
polydisperse hard spheres, J. Chem. Phys. \textbf{138}, 12A536 (2013).

\bibitem{kawasaki2007correlation} T. Kawasaki, T. Araki, and H. Tanaka,
Correlation between dynamic heterogeneity and medium-range order in
two-dimensional glass-forming liquids, Phys. Rev. Lett. \textbf{99}, 215701
(2007).

\bibitem{russo2015assessing} J. Russo and H. Tanaka, Assessing the role of
static length scales behind glassy dynamics in polydisperse hard disks,
Proc. Natl. Acad. Sci. U.S.A. \textbf{112}, 6920 (2015).

\bibitem{Adams} D. J. Adams, Chemical potential of hard-sphere fluids by
Monte Carlo methods, Mol. Phys. \textbf{28}, 1252 (1974).

\bibitem{Andrews} F. C. Andrews, Simple approach to the equilibrium
statistical mechanics of the hard sphere fluid, J. Chem. Phys. \textbf{62},
272 (1975).

\bibitem{CMP2023} V. M. Pergamenshchik, Incorporation of the intensive and
extensive entropy contributions in the disk intersection theory of a hard
disk system, Cond. Mat. Phys. \textbf{26}, 33501 (2023).

\bibitem{BaxterReview} R. J. Baxter, Solving models in statistical
mechanics,\ Adv. Studies in Pure Math. \textbf{19}, 95 (1989).

\bibitem{Baxter} R. J. Baxter,\emph{\ }\textit{Exactly solved models in
statistical mechanics} (Academic Press, Great Britain, 1989).

\bibitem{Analyticaltheory} This is a usual requisite of solutions of
many-body problems: solving an infinite-dimensional PF consists in
identification of a finite number of some quantities that enable one to get
the PF \cite{Baxter}. In the majority of solvable models these quantities
(e.g., eigenvalues of eigenfunctions of certain transfer matrix) can be
obtained only numerically.

\bibitem{McQuarrie} D. A. McQuarrie, \textit{Statistical Mechanics}
(University Science Books, 2000).

\bibitem{ExperColloids} A. L. Thorneywork, J. L. Abbott, D. G. A. L. Aarts,
and R. P. A. Dullens, Two-dimensional melting of a colloidal hard spheres,
Phys. Rev. Lett., \textbf{118}, 158001 (2017).

\bibitem{KTHNY} J. M. Kosterlitz, Kosterlitz-Thouless physics: a review of
key issues Rep. Prog. Phys. \textbf{79}, 026001 (2016).

\bibitem{ref44} A. Huerta, D. Henderson, and A. Trokhymchuk, Freezing of
two-dimensional hard disks, Phys. Rev. E \textbf{74}, 061106 (2006).

\bibitem{KT1973} J. M. Kosterlitz and D. J. Thouless, Ordering,
metastability and phase transitions in two-dimensional systems, J. Phys. C 
\textbf{6}, 1181 (1973).

\bibitem{Trokhymchuk2025} A. Trokhymchuk, A. Huerta, and T. Bryk, Bimodality
of local structural ordering in extremely confined hard disks, J. Chem.
Phys., \textbf{162}, 184504 (2025).

\bibitem{Psi6} D. R. Nelson and B. I. Halperin, Dislocation-mediated melting
in two dimensions, Phys. Rev. B \textbf{19}, 2457 (1979).

\bibitem{ref36} J. M. Kosterlitz and D. J. Thouless, Long range order and
metastability in two dimensional solids and superfluids, J. Phys. C \textbf{5%
}, L124 (1972).

\bibitem{ref38} K. W. Kratky, The area of intersection of n equal circular
disks, J. Phys. A \textbf{11}, 619 (1978).

\bibitem{ref39} K. W. Kratky, Intersecting disks (and spheres) and
statistical mechanics. I. Mathematical basis, J. Stat. Phys. \textbf{25},
619 (1981).
\end{thebibliography}
\end{document}